\documentstyle[11pt]{article}
\input epsf
\def\thefootnote{\fnsymbol{footnote}}

\newcommand{\rll}{\rule[-0.3cm]{0cm}{0.9cm}}
\newcommand{\rlll}{\rule[-0.3cm]{0cm}{1.2cm}}

\newcommand{\sps}{\hspace{3mm}}

\newcommand{\nn}{\nonumber}

\newcommand{\de}{\partial}
\newcommand{\ri}{\right}
\newcommand{\lf}{\left}

\newcommand{\ep}{\varepsilon}
\newcommand{\eq}{\begin{equation}}
\newcommand{\en}{\end{equation}}
\newcommand{\bea}{\begin{eqnarray}}
\newcommand{\eea}{\end{eqnarray}}
\newcommand{\acc}{\nn\\[4mm]}
\newcommand{\ba}{\begin{array}}
\newcommand{\ea}{\end{array}}
\newcommand{\ds}{\displaystyle}
\newcommand{\II}{\hbox{{\rm l{\hbox to 1.5pt{\hss\rm l}}}}}
\newcommand{\ZZ}{{\hbox{$\sf\textstyle Z\kern-0.4em Z$}}}
\newcommand{\Y}{\Upsilon}
\newcommand{\CM}{{\cal{M}}}

\newcommand{\CX}{{\cal{X}}}
\newcommand{\CZ}{{\cal{Z}}}

\newcommand{\virg}{~,}

\newcommand{\resection}[1]{\setcounter{equation}{0}\section{#1}}

\newcommand{\half}{\frac{1}{2}}

\newcommand{\h}{{\cal H}}
\newcommand{\HH}{{\sf h}}
\newcommand{\NP}[1]{Nucl.\ Phys.\ {\bf #1}}
\newcommand{\PL}[1]{Phys.\ Lett.\ {\bf #1}}

\newcommand{\CMP}[1]{Comm.\ Math.\ Phys.\ {\bf #1}}
\newcommand{\PR}[1]{Phys.\ Rev.\ {\bf #1}}
\newcommand{\PRL}[1]{Phys.\ Rev.\ Lett.\ {\bf #1}}
\newcommand{\PTP}[1]{Prog.\ Theor.\ Phys.\ {\bf #1}}
\newcommand{\MPL}[1]{Mod.\ Phys.\ Lett.\ {\bf #1}}
\newcommand{\IJMP}[1]{Int.\ J.\ Mod.\ Phys.\ {\bf #1}}
\newcommand{\JETP}[1]{Sov.\ Phys.\ JETP {\bf #1}}

\newcommand{\ABZ}{A.B.Zamolodchikov}
\newcommand{\AlBZ}{Al.B.Zamolodchikov}
\newcommand{\ZN}{\ZZ_N}
\newcommand{\e}{{\rm e}}
\newcommand{\ddt}{{d\over d\theta}}
\newcommand{\AP}[1]{Ann.\ Phys.\ {\bf #1}}
\newcommand{\JMP}[1]{J.\ Math.\ Phys.\ {\bf #1}}
\newcommand{\JSP}[1]{J.\ Stat.\ Phys.\ {\bf #1}}
\newcommand{\JP}[1]{J.\ Phys.\ {\bf #1}}
\newcommand{\JPh}[1]{J.\ Physique.\ {\bf #1}}
\newcommand{\ubl}[1]{\{#1\}}
\newcommand{\usbl}[1]{\left(#1\right)}
\newcommand{\iint}{\int^{\infty}_{-\infty}}
\newcommand{\VEV}[1]{\langle #1\rangle}
\newcommand{\Epsilon}{{\cal E}}
\newcommand{\MR}{M\! R}
\newcommand{\wtilde}{\widetilde}
\newcommand{\sscr}{\scriptscriptstyle}
\newcommand{\gam}[1]{\gamma\lf({\scriptstyle{#1}}\ri)}
\newcommand{\Gam}[1]{\Gamma\lf({\scriptstyle{#1}}\ri)}
\newcommand{\Cam}[1]{C^{\scriptscriptstyle{(#1)}}}
\newcommand{\CTam}[1]{\wtilde C^{\scriptscriptstyle{(#1)}}}
\newcommand{\epL}{\ep^{\scriptscriptstyle{L}}}
\newcommand{\epR}{\ep^{\scriptscriptstyle{R}}}
\newcommand{\LL}{L^{\scriptscriptstyle{L}}}
\newcommand{\LR}{L^{\scriptscriptstyle{R}}}
\newcommand{\YL}{Y^{\scriptscriptstyle{L}}}
\newcommand{\YR}{Y^{\scriptscriptstyle{R}}}
\newcommand{\Flog}{F_{(\log)}^{\scriptscriptstyle{(UV)}}}
\newcommand{\widetable}{\renewcommand{\arraystretch}{1.65}}
\newcommand{\fracs}[2]{{\scriptstyle\frac{#1}{#2}}}
\newcommand{\fract}[2]{{\textstyle\frac{#1}{#2}}}
\newcommand{\halfs}{\fracs{1}{2}}
\newcommand{\halft}{\fract{1}{2}}
\newcommand{\dth}{\partial_{\theta}}
\newcommand{\m}{\phantom{-}}
\newcommand{\vs}{\vskip 4pt}
\newcommand{\opnup}[1]{\renewcommand{\\}{\\[50 pt]}}
\renewcommand{\bar}{\overline}
\begin{document}
\begin{titlepage}
\vskip 0.5cm
\begin{flushright}
DTP-95/81 \\
January 1996 \\
hep-th/9601123
\end{flushright}
\vskip 1.5cm
\begin{center}
{\Large {\bf Massive and massless phases in self-dual}} \\[5pt]
{\Large {\bf $\ZZ_N$ spin models: some exact results from}  }\\[5pt]
{\Large {\bf the thermodynamic Bethe ansatz}  }
\end{center}
\vskip 0.9cm
\centerline{Patrick Dorey, Roberto Tateo and Kevin\,E.\,Thompson}
\vskip 0.6cm
\centerline{\sl Department of Mathematical Sciences,  
}
\centerline{\sl  University of Durham, Durham DH1 3LE, 
England\,\footnote{e-mail: {\tt P.E.Dorey,
Roberto.Tateo, K.E.Thompson~~@durham.ac.uk}} }
\vskip 1cm
\begin{abstract}
\vskip0.2cm
\noindent

The generally accepted phase diagrams for the discrete $\ZN$ spin 
models in two dimensions imply the existence of certain renormalisation
group flows, both between conformal field theories and into a massive 
phase. Integral equations are proposed to describe these flows, and 
some properties of their solutions are discussed. The infrared behaviour
in massless and massive directions is analysed in detail, and the
techniques used are applied to a number of other models.

\end{abstract}
\end{titlepage}

\setcounter{footnote}{0}
\def\thefootnote{\arabic{footnote}}

\resection{ Introduction }

A planar vector model in which an $O(2)$ symmetry has been broken down
to a $\ZN$ subgroup can be defined by the reduced
Hamiltonian
\eq
\h [\{\theta_i\}]=\sum_{\langle ij\rangle}V(\theta_i-\theta_j)-\HH\sum_i\cos
N\theta_i~.
\label{hamiltonian}
\en
The pairs $\langle ij\rangle$ are nearest-neighbour sites of the
two-dimensional lattice on which the angular spins $\theta_i$ 
live; these interact via a $2\pi$-periodic, even function $V(\theta)$.
The strength of the explicit symmetry-breaking is governed by 
the coupling $\HH$. 

When $\HH{=}0$, continuous symmetry is regained and this 
forbids the formation of an ordered phase, no matter how low the
temperature $T$ becomes. Instead, 
the model undergoes a Kosterlitz-Thouless transition 
as $T$ decreases through some critical
value $T_c$. Below this temperature, the completely disordered 
high-temperature phase, in which the typical spin configuration is
dominated by vortices, is replaced by a massless phase in which the 
vortices are suppressed and correlations decay algebraically. 
For $\HH\neq 0$, this picture must 
change: since residual symmetry is discrete there is no obstruction
to its spontaneous breaking at low enough temperatures, and a
rigorous result of Fr\"ohlich and Lieb asserts that this does indeed
happen~\cite{FLa}.
Despite this fact, for $N\geq 5$ the massless phase is not
completely lost: Jos\'e {\it et al}~\cite{JKKNa} showed that the $\ZN$
perturbation is irrelevant above a temperature
$T_{c'}$ which, for $N\geq 5$, lies {\it below} $T_c$. A
small non-zero value of $\HH$ cannot change the $\HH{=}0$ critical behaviour
until $T$ falls below $T_{c'}$, and so the Kosterlitz-Thouless phase
survives as an intermediate stage between massive high- and low-
temperature regimes.

Another situation amenable to analysis is the limit
$\HH\rightarrow\infty$: this serves to pin the spins to $N$ discrete
values, and results in a `pure' $\ZN$ model. Universality would
suggest the continuing validity of the picture just outlined, and this
expectation was confirmed by Elitzur {\it et al} in 1979~\cite{EPSa}. 
However, both this and the earlier work of Jos\'e {\it et al}
concentrated on the Villain form~\cite{Va} of the interaction $V(\theta)$,
and it is important to know how much influence this choice has on the
nature of the phases and transitions exhibited by the model.
For the pure $\ZN$ case,
it is feasible to investigate this question directly, since the 
full phase space is finite ($[N/2]$) dimensional: a particular
system is specified by the $N$ reduced
energy differences $V_r= V(2\pi r/N){-}V(0)$,
subject to the condition $V_r=V_{N-r}$.
The general features of the resulting phase diagrams have been the subject 
of much attention since the early 1980's,
motivated both by the intrinsic interest of the problem and by the
analogies between such two-dimensional spin systems and certain
four-dimensional gauge theories~\cite{DMSa}.

For each $N$, a duality transformation can be defined as 
follows~\cite{WWa}. First parametrise the couplings as 
$x_r=\exp(-V_r)$. Then the
summation over the spins $\theta_i$ can be replaced by one over
dual spins $\wtilde\theta_{{\sscr\tilde\imath}}$ defined on the sites
$\tilde\imath$ of the dual lattice, so long as the set of
couplings $\{x_r\}$
is simultaneously replaced by the dual set $\{\wtilde x_r\}$:
\[
\wtilde x_r=\lf(1+\sum_{s=1}^{N-1}x_s e^{2\pi\imath rs/N}\ri)\!\Bigg/\!
\lf(1+\sum_{s=1}^{N-1}x_s\ri)~.
\]
The effect 
is to exchange the low and high temperature fixed points 
($x_r{\equiv}0$ and $x_r{\equiv}1$ respectively), while leaving an
$[N/4]$-dimensional hyperplane invariant -- systems on this
hyperplane are said to be self-dual.

Setting $\sigma^k_i=\e^{\imath k\theta_i}$ and $\mu^l_{\tilde\imath}
=\e^{\imath l\tilde\theta_{\tilde\imath}}$, the various phases can be
characterised by the order parameters $\langle\sigma^k\rangle$ and
their duals, the `disorder parameters' $\langle\mu^l\rangle$.
In a completely ordered phase, 
$\langle\sigma^k\rangle\neq 0$, $\langle\mu^l\rangle=0$;
in a completely disordered phase $\langle\sigma^k\rangle=0$, 
$\langle\mu^l\rangle\neq 0$; while both 
vanish in a Kosterlitz-Thouless phase.
In addition, if $N$ is not prime there are `partially ordered'
phases where expectation values of certain powers
$\langle\sigma^k\rangle$ or $\langle\mu^l\rangle$ 
are nonzero even though others 
vanish~\cite{DRa}.
However such phases, and the corresponding partial-ordering
transitions, are found away from the region that will be of interest
below. 

\begin{figure}[h]
\hspace{2.7cm}
\epsfxsize=190pt
\epsfysize=190pt
\epsfbox{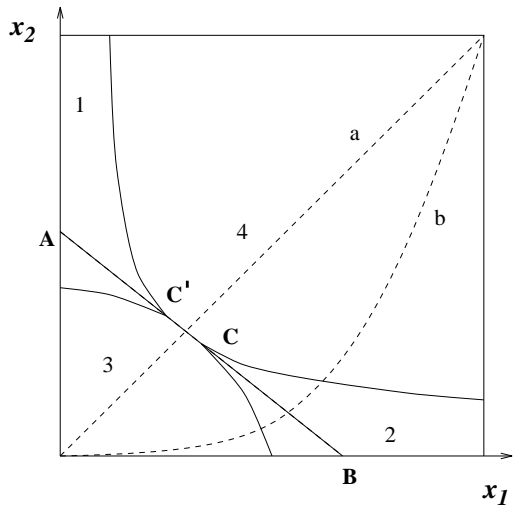}
\caption{ Phases for the general $\ZZ_5$
model: the Kosterlitz-Thouless regions $1$ and $2$ are massless, while
regions $3$ (ordered) and $4$ (completely disordered) are massive. The
line $AB$ is self-dual; $a$ labels the line of Potts models, and $b$ is the
Villain line.}
\label{phsfig} 
\end{figure}

The way that the three phases which {\it are} important for this
paper fit together can be illustrated with the case 
$N{=}5$, shown in figure~\ref{phsfig}. The transition from the ordered to
the disordered phase can happen in one of two ways: either via an
intermediate massless phase, as along the Villain line $(b)$,
or else directly through a single first-order transition, as
exemplified by the behaviour along the Potts line $(a)$.
Clearly, the points $C$ and $C'$ on the figure
are rather special: they mark the
bifurcations of the line of first-order transitions into pairs of
Kosterlitz-Thouless transitions, and the beginnings of the
massless regions. These points are thought to be special for other
reasons too. In 1982 Fateev and Zamolodchikov identified particular 
sets of self-dual Boltzman weights
for the general $\ZN$ model which satisfy the
star-triangle relations, and conjectured that they were precisely
the critical points $C$ and $C'$ for $N{=}5$, or their
generalisations to higher $N$~\cite{FZa}. Subsequently, in
1985, they found for each $N$ a $\ZN$-symmetric conformal field theory
that was a natural candidate for the field-theory limit of these
same special points~\cite{ZFa}.

To verify these claims, two distinct
questions must be addressed. First, the criticality of the
Fateev-Zamolodchikov Boltzman weights should be established and their
critical exponents matched with those of the corresponding
$\ZN$-symmetric conformal field theory: this has now been done, in 
both numerical~\cite{Aa} and analytical~\cite{JMOa} studies, with 
complete agreement
with expectations. However, on its own this information is
not enough to locate the models relative to the opening of the
first-order transition into a 
massless phase. To answer this second question, neighbouring regions
of the phase diagram must be examined. Alcaraz~\cite{Aa} found
numerical results consistent with the conjecture for the $\ZZ_5$
case, but beyond this, direct evidence from
the lattice model is hard to come by. But the question
can also be studied in the continuum, using the
techniques pioneered by \ABZ\ in his work on
perturbed conformal field theories~\cite{Za}. If the system
described by the $\ZN$-symmetric conformal field theory does indeed
lie on the border of a Kosterlitz-Thouless phase, then a suitable 
perturbation will
shift it either into a massive phase, or else into the massless
Kosterlitz-Thouless region. 
Correspondingly, the conformal theory should admit perturbations
$S_{\rm CFT}\rightarrow S_{\rm CFT}+\lambda\int\!\epsilon\,d^2x$ which
provoke both massive and massless renormalisation group flows,
according to the sign of the coupling $\lambda$. The perturbing
operator $\epsilon$ should be $\ZN$-symmetric, since the 
initial shift of the Hamiltonian is within the phase diagram of
the $\ZN$ models; and taking $\epsilon$ to be self-dual will ensure
that any massive direction of the perturbation is onto the surface
of first-order transitions. The massless
direction, shifting the model into a Kosterlitz-Thouless phase,
should lead to a flow from $c=2(N{-}1)/(N{+}2)$, the central charge
of the $\ZN$-symmetric conformal field theory, to $c=1$, the central
charge of the various one-component Gaussian models.

If $\epsilon$ can be chosen so that the perturbed theory is
integrable, then past experience of a method known as the
Thermodynamic Bethe Ansatz (TBA) suggests that there should exist a
set of integral equations encoding the {\it exact} flow of the 
ground-state energy $E(R)$ of the system on a 
cylinder of circumference $R$, from $R{=}0$ 
all the way through to $R{=}\infty$. Equations of this type were
initially derived for theories for which a conjectured exact
S-matrix was already available~\cite{Zb,KMa}, but it has since
proved possible to make educated guesses for equations to describe many
ground-state energy flows even in the absence of direct 
derivations~\cite{Zc,Zd}.

The purpose of this paper is to
propose just such sets of equations, pertinent to
self-dual flows from the
$\ZN$-symmetric conformal field theory into either a massive or a
Kosterlitz-Thouless phase. 
The systems themselves are given in the next section, while section~3
shows that at small $R$ they
match a suitable
perturbation already shown by Fateev to be integrable~\cite{Fa}.
The discussion then turns to the large~$R$ behaviour, starting with
the massless direction. In section~4.1, this is shown to be compatible
with a self-dual flow into the Kosterlitz-Thouless region.
We were able to extract some exact infrared information directly
from the TBA, and this is explained in section~4.2.
The technique is applied in a number of contexts,
with results that reflect the connections between
the various models associated with $c{=}1$ conformal field
theories. In the massive direction, described in section~4.3,
a detailed examination of the asymptotics allows us to conjecture a
pattern of massive kinks linking $N{+}1$ degenerate
vacua, as expected on a surface of first-order transitions where
phase coexistence occurs.
The concluding section~5 outlines some
questions that might merit further study.

A number of the more technical details have been relegated to a series 
of appendices:
the integral kernels, Y-systems and
dilogarithm sum rules used in the main text
can be found in appendices A, B and C, while numerical results
for the ultraviolet asymptotics are reported in  
appendix~D.

\resection{TBA equations}

In this section, the equations will be introduced, leaving
a detailed study of their properties until later. Some 
are already in the
literature~\cite{Ra} -- \cite{RTVa}\thinspace. Since these 
motivate the subsequent proposals, the section
begins with a review of this material.

\subsection{Known cases}

The initial goal is to find sets of equations consistent with
movement from 
$c=2(N{-}1)/(N{+}2)$ to $c=1$. For $N$ even such behaviour has been seen
before. Each $W_{SO(m)}$ series of coset conformal theories allows 
for interpolating flows, analogous to the
$\phi_{13}$-induced flows in the minimal series. The final step,
\eq
{SO(m)^{(2)}\times SO(m)^{(1)}\over SO(m)^{(3)}}
+\lambda\phi_{\rm 1,1,Adj} \quad\rightarrow\quad
{SO(m)^{(1)}\times SO(m)^{(1)}\over SO(m)^{(2)}}
\label{cosets}
\en
involves a change of central charge
$2(2m{-}1)/(2m{+}2)\rightarrow 1$,
which fits the bill for $N{=}2m$. This was noted by Fateev~\cite{Fa}, 
and our only contribution here will be to exploit
the TBA systems that have since been proposed:
$d_{n+1}$-related for $m{=}2n{+}2$~\cite{Ra} 
and $b_n$-related for $m{=}2n{+}1$~\cite{KNa,Ta}
($n$ has been picked in this way for later convenience: in
particular, the choice means that the associated sine-Gordon models
always have $n{-}1$ breathers). 
The lack of an explicit $\ZZ_{2m}$ symmetry in the  
models~(\ref{cosets}) is not necessarily a problem, particularly as 
attention is being restricted to the flow of the ground-state energy.
The real test will come later, when a more detailed look is taken at
the predicted asymptotic behaviours of $E(R)$ at small and large values 
of~$R$.

Thus, for $N{=} 4n{+}4$ we conjecture that the ground-state
energy will be as in ref.~\cite{Ra}:
\eq
RE(R)=-{1\over 2\pi}\!\!\!\sum^{{}\atop n{+}1}_{i=1\atop\alpha=1,2}
 \iint\!d\theta\,\nu_i^{(\alpha)}(\theta)L^{(1)}_i(\theta)
\label{Eziv}
\en
where, both here and subsequently,
$$
L^{(\alpha)}_i(\theta)=\log(1+e^{-\ep_i^{(\alpha)}(\theta)})~.
$$
The functions $\ep^{\sscr(\alpha)}_i(\theta)$, $i=1\dots n{+}1$,
are to be found as the solutions to
the following set of coupled integral equations:
\eq
\ep^{(\alpha)}_i(\theta)=\nu^{(\alpha)}_i(\theta)-
\!\!\sum^{n{+}1}_{j=1}
\!\lf[\phi_{ij}{*}L^{(\alpha)}_j(\theta)
-\psi_{ij}{*}L^{(\wtilde\alpha)}_j(\theta)\ri]\, ~~~(\alpha=1,2)\,,
\label{TBAziv}
\en
with $\wtilde\alpha=3{-}\alpha$, and $*$ denoting the convolution 
$$
f{*}g(\theta)={1\over 2\pi}\iint\!f(\theta')g(\theta-\theta')d\theta'~.
$$
The kernels $\phi_{ij}$ and $\psi_{ij}$ can be extracted from 
equation~(\ref{kernels}) of appendix~A
on setting $h{=}2n$. They can equally be defined
using the $d_{n{+}1}$ Toda S-matrices, though this fact will only
briefly be relevant below.
The `energy terms' $\nu^{(\alpha)}_i(\theta)$ are
\eq
\nu^{(1)}_i(\theta)=\nu^{(\wtilde 1)}_i(-\theta)
   =\half M_iRe^{\theta}\qquad (i=1\dots n{+}1)\,,
\label{energyterms}
\en
and the numbers $M_i$ are given by (\ref{bmasses}),(\ref{bbmasses}), again 
with $h{=}2n$. The full system has a $\ZZ_2$ symmetry under
$\alpha\rightarrow\wtilde\alpha$, $\theta\rightarrow-\theta$.
This transformation has a simple interpretation, as follows.
The pseudoenergies $\ep^{\sscr(1)}_i$
and $\ep^{\sscr(\wtilde 1)}_i$ can be associated with right and left
moving massless `particles'~\cite{Zd}, with S-matrix elements
\[
S^{LL}_{ij}(\theta)=S^{RR}_{ij}(\theta)=S_{ij}(\theta)\quad,\quad
S^{RL}_{ij}(\theta)=S^{LR}_{ij}(\theta)=\lf(T_{ij}(\theta)\ri)^{-1}
\]
(with $S_{ij}$ and $T_{ij}$ as in appendix~A). The
$\ZZ_2$ transformation reverses their spatial momenta 
(by sending $\theta$ to $-\theta$) and swaps right
and left movers; it therefore implements parity. 
In the rest of the paper, this sort of language will often be used,
even though the general problem of associating S-matrices to the flows
is being left for future work.

Were the theory conformal, $E(R)$ would be given in terms of the
central charge $c$ by the relation
$E(R)=-\pi c/6R$~\cite{BCNAa}. For a non-conformal theory an
`effective central charge' $c(R)$ can still be defined, as
$$E(R)=-{\pi\over 6R}c(R)\,,$$
after which the ultraviolet and infrared central charges
can be extracted as the $R{\rightarrow}0$ and
$R{\rightarrow}\infty$ limits of $c(R)$.\footnote{Strictly speaking,
any bulk terms must also be subtracted before the infrared result is
valid. The TBA does this automatically.}
These limits follow from~(\ref{Eziv}) and~(\ref{TBAziv}) 
via by-now standard manipulations,
and are indeed as claimed.
For the moment the only point
to note is the way that the $R{\rightarrow}\infty$ limit
works: $L^{(2)}_i(\theta)$ becomes vanishingly
small at any values of $\theta$ for which 
$L^{(1)}_i(\theta)$ is non-zero, and vice versa, 
the two halves of (\ref{TBAziv}) decouple, and the equations become
\eq
\ep^{(\alpha)}_i(\theta)=\nu^{(\alpha)}_i(\theta)-
\sum^{n{+}1}_{j=1}\!\phi_{ij}{*}L^{(\alpha)}_j(\theta)
\, ,~~~(\alpha=1,2).
\label{TBAzivd}
\en
Noting that the $R$-dependence can be removed by 
shifts in $\theta$ of $\pm\log R$, these `kink systems' can be recognised as 
two copies of the scale-independent
TBA equations associated with the minimal $d_{n+1}$-related
S-matrices, and also, more suggestively, with the 
sine-Gordon model at the reflectionless points
\eq
\beta^2={8\pi\over n{+}1}\equiv{32\pi\over N}~.
\label{SGcouplingiv}
\en

For $N{=}4n{+}2$ the story is much the same. The underlying
massless scattering theory is not expected to be diagonal, and as a
consequence the TBA system includes some auxiliary functions, in
addition to those $\ep^{\sscr(\alpha)}_i$ which appear explicitly in the
formula for the ground-state energy. These functions, often
referred to as `magnonic' pseudoenergies, make themselves felt
via their appearance in the coupled set of integral equations that all
the pseudoenergies must satisfy together. For the ($b_n$-related) case
in hand, three such functions must be introduced. It will be
convenient to label them as $\ep^{\sscr(0)}_n$, $\ep^{\sscr(2)}_n$ and
$\ep^{\sscr(4)}_n$, with the remaining non-magnonic pseudoenergies 
being $\ep^{\sscr(1)}_i$ and $\ep^{\sscr(3)}_i$, $i=1\dots n$.
(This notation is in line with the idea that it is
particle number $n$ that should be blamed for the magnonic 
terms in the equations, $n$ being the label of the
soliton in the related sine-Gordon model, while $1\dots n{-}1$
label the diagonally-scattering sine-Gordon breathers.)
The appropriate
$b_n$-related TBA system can then be written as
\eq
RE(R)=-{1\over 2\pi}\!\sum^{{}\atop n}_{i=1\atop\alpha=1,3}
 \iint\!d\theta\,\nu_i^{(\alpha)}(\theta)L^{(1)}_i(\theta)
\label{ENzbn}
\en
with
\bea\ds
\ep^{(\alpha)}_i(\theta)&=&\nu^{(\alpha)}_i(\theta)-
\!\sum^n_{j=1}
\lf[\phi_{ij}{*}L^{(\alpha)}_j(\theta)
-\psi_{ij}{*}L^{(\wtilde\alpha)}_j(\theta)\ri]\nn\\
&&~~~~{}-\delta_{i,n}\!\sum^4_{\beta=0}l_{\alpha\beta}^{[a_5]}
           \phi_1{*}L^{(\beta)}_n(\theta)\, ,~~~~~(\alpha=1,3)\,;\nn\\
\ep^{(\alpha)}_n(\theta)&=&-
\sum^4_{\beta=0}l^{[a_5]}_{\alpha\beta}
     \phi_1{*}L^{(\beta)}_n(\theta)\,,~~~~~\quad(\alpha=0,2,4)\,.
\label{TBAzbn} \\ \nn
\eea
\vskip -4pt\noindent
Setting $\wtilde\alpha=4{-}\alpha$, the energy terms are again given by
equation~(\ref{energyterms}). The masses $M_i$, and the kernels
$\phi_{ij}$, $\psi_{ij}$ and $\phi_1$, can be found 
in appendix~A, with $h=2n{-}1=N/2{-}2$.
Parity symmetry is implemented as
$\alpha\rightarrow\wtilde\alpha$, $\theta\rightarrow -\theta$.

In the $R{\rightarrow}\infty$ limit, the interesting behaviour is
confined to kink systems near $\theta=\pm\log R$, and again the
equations separate, into one set in which the index $\alpha$ takes the
values $0,1,2$ and a second in which it takes the values $2,3,4$.
As before, the resulting equations could
have been found by examining the ultraviolet limit of a
sine-Gordon TBA system, this time at the coupling
\eq
\beta^2={8\pi\over n+1/2}\equiv{32\pi\over N}~.
\label{SGCouplingbn}
\en

Having observed this connection,
we can hope to use it in reverse to construct new TBA systems for
the remaining ($N$ odd) cases, using as an input the known sine-Gordon
TBA systems at yet further values of the
coupling $\beta$. The question is, just which values? Looking at
equations (\ref{SGcouplingiv}) and (\ref{SGCouplingbn}), a natural
guess comes to mind, and to confirm it we now recall the one other
case where an exact conjecture for the ground-state energy 
has previously appeared, namely
the $N{=}5$ flow from $c{=}8/7$ to $c{=}1$. This was
found by Ravanini {\it et al\/} in the course of an exhaustive study
of TBA systems of a particular type, and reads~\cite{RTVa}:
\eq
RE(R)=- {1\over 2\pi}\iint\!d\theta\lf[\nu^{(1)}(\theta)L^{(1)}(\theta)+
              \nu^{(6)}(\theta)L^{(6)}(\theta) \ri]~,
\en
where the functions 
$\ep^{(1)}(\theta)$ and $\ep^{(6)}(\theta)$ appearing in
$L^{(1)}$ and $L^{(6)}$ are coupled,
together with four auxiliary functions $\ep^{(2)}\dots\ep^{(5)}$,
in the following system of equations:
\eq
\ep^{(\alpha)}(\theta)=\nu^{(\alpha)}(\theta)-
\!\sum^6_{\beta=1}
l_{\alpha\beta}^{[e_6]}
           \phi{*}L^{(\beta)}(\theta)\, ~~~~~(\alpha=1\dots 6)~.
\label{TBAzV}
\en
Here $\nu^{(\alpha)}(\theta)= \halft\MR e^{\theta} \delta_{1\alpha}
+\halft\MR e^{-\theta} \delta_{6\alpha}\,$, $\phi(\theta)=1/\!\cosh\theta$,
and $l^{[e_6]}_{\alpha\beta}$
is the incidence matrix of the $e_6$ Dynkin diagram, labelled so that
$1$ and $6$ are the two extremal nodes. The $\ZZ_2$ symmetry of this
diagram, combined with $\theta\rightarrow -\theta$, exchanges left and
right movers and thus implements parity.
The
$R\rightarrow\infty$ limit removes all trace of $\ep^{(6)}$ from 
the equations satisfied by $\ep^{(1)}$, and vice versa;
thus in this limit
\eq
\ep^{(\alpha)}(\theta)=\nu^{(\alpha)}(\theta)-
\!\sum^5_{\beta=1}
l_{\alpha\beta}^{[d_5]}
           \phi{*}L^{(\beta)}(\theta)~~~~~(\alpha=1\dots 5),
\label{TBAzzV}
\en
with $l^{[d_5]}_{\alpha\beta}$ the
incidence matrix of the $d_5$ Dynkin diagram. (An analogous equation
determines the infrared form of $\ep^{\sscr(6)}(\theta)$.)
Just as in the earlier 
cases, the final TBA system could equally have emerged in a
discussion of the 
thermodynamics of the sine-Gordon model. This time, the coupling to
take is $\beta^2=32\pi/5$~\cite{FIa}, precisely the value 
predicted by equations (\ref{SGcouplingiv}) and~(\ref{SGCouplingbn}).

\subsection{New massless TBA systems}
By now it is natural to suppose that a system of equations
appropriate to the
$N{=}7$ flow will have something to do with the sine-Gordon TBA at
$\beta^2=32\pi/7$. In its ultraviolet limit, this latter system
becomes 
\bea\ds
\ep^{(1)}(\theta)
&=&\nu^{(1)}(\theta)-\phi_3{*}(L^{(4)}(\theta)+L^{(5)}(\theta))
  -\phi_4{*}L^{(2)}(\theta)-\phi_5{*}L^{(3)}(\theta)\nn\\
\ep^{(2)}(\theta)
&=&\phi_2{*}(K^{(3)}(\theta)-L^{(1)}(\theta))\nn\\
\ep^{(3)}(\theta)
&=&\phi_2{*}(K^{(2)}(\theta)+K^{(4)}(\theta)+K^{(5)}(\theta))~~;~~~
 \ep^{(4)}(\theta)=\phi_2{*}K^{(3)}(\theta)\nn\\
\ep^{(5)}(\theta)&=&\phi_2{*}K^{(3)}(\theta)\nn\\
\label{TBAsgvii}
\eea
\vskip -12pt\noindent
where
$$
L^{(\alpha)}=\log(1+e^{-\ep^{(\alpha)}})~,~~
K^{(\alpha)}=\log(1+e^{\ep^{(\alpha)}})~,~~\nu^{(1)}(\theta)=\half
\MR e^{\theta}~,
$$
and the kernels are given by equation~(\ref{magkernels}) with
$h=3/2$. (One way to obtain this system is to Fourier transform the
Y-system given in ref.~\cite{Tb}, divide through by appropriate
hyperbolic cosines, and then transform back.)

The scale $R$ enters into these equations in a
trivial way, and can be removed by a shift in $\theta$. This is as it
should be, since the ultraviolet limit has already been taken.
The scale-dependent sine-Gordon TBA could be recovered on redefining
$\nu^{(1)}(\theta)=\MR\cosh\theta$, 
a man\oe uvre that leaves the $R{\rightarrow}0$
limit unchanged but forces 
$c_{SG}(\infty){=}0$, as appropriate for a massive theory. However,
here the idea is different -- we
want~(\ref{TBAsgvii}) to describe the infrared destination of some
massless flow, rather than the ultraviolet limit of a
massive one. To this end, the scale-invariant system, thought
of as describing only
leftmoving particles, should be augmented
with an extra piece to describe the
right-movers. This new piece will re-introduce a scale dependence, but
in such a way that the system~(\ref{TBAsgvii}) is
recovered as $R\rightarrow\infty$. Furthermore, the
augmented system must treat left and right movers on an equal
footing, with a $\ZZ_2$ symmetry exchanging them. Remarkably, 
the system~(\ref{TBAsgvii}) is such that both of these
requirements can be satisfied in a non-trivial way. 
On noticing the $\ZZ_2$ symmetry already exhibited by the mutual
couplings of the magnonic pseudoenergies $\ep^{(2)}\dots\ep^{(5)}$,
it is natural to extend this to the full system with the addition of a
single right-moving pseudoenergy $\ep^{(6)}(\theta)$:
\bea\ds
\ep^{(1)}(\theta)
&=&\nu^{(1)}(\theta)-\phi_3{*}(L^{(4)}(\theta)+L^{(5)}(\theta))
-\phi_4{*}L^{(2)}(\theta)-\phi_5{*}L^{(3)}(\theta)\nn\\
\ep^{(2)}(\theta)&=&\phi_2{*}(K^{(3)}(\theta)-L^{(1)}(\theta))\nn\\
\ep^{(3)}(\theta)
&=&\phi_2{*}(K^{(2)}(\theta)+K^{(4)}(\theta)+K^{(5)}(\theta))~~;~~~
 \ep^{(4)}(\theta)=\phi_2{*}K^{(3)}(\theta)\nn\\
\ep^{(5)}(\theta)&=&\phi_2{*}(K^{(3)}(\theta)-L^{(6)}(\theta))\nn\\
\ep^{(6)}(\theta)
&=&\nu^{(6)}(\theta)-\phi_3{*}(L^{(4)}(\theta)+L^{(2)}(\theta))
-\phi_4{*}L^{(5)}(\theta)-\phi_5{*}L^{(3)}(\theta)\nn\\
\label{TBAzvii}
\eea
Taking $\nu^{(6)}=\half\MR e^{-\theta}$ gives the desired
left-right symmetry, under 
$\ep^{(1)}{\leftrightarrow}\ep^{(6)}$,
$\ep^{(2)}{\leftrightarrow}\ep^{(5)}$, $\theta{\rightarrow} {-\theta}$.
If the ground state energy is 
\eq
RE(R)=-{\pi\over 6} c(R)=
  -{1\over 2\pi}\iint\!d\theta\lf[\nu^{(1)}(\theta)L^{(1)}(\theta)+
              \nu^{(6)}(\theta)L^{(6)}(\theta) \ri]~,
\label{ENzvii}
\en
then the result $c(\infty){=}1$ follows, as intended, from the result
$c_{SG}(0){=}1$ for the sine-Gordon model -- the determining 
equations reduce to~(\ref{TBAsgvii}) in both cases.
More interesting is the $R{\rightarrow}0$ limit of the extended 
system~(\ref{TBAzvii}), 
and this confirms that the discussion thus far has
been on the right track. Using the dilogarithm sum rules
given in appendix C,
it can be checked that $c(0)=4/3$, exactly as required.

Once this case has been understood, the general procedure is
clear. Take the sine-Gordon TBA at $\beta^2{=}32\pi/N$, and look for
a $\ZZ_2$ symmetry of its magnonic part, which is not a symmetry
of the full system. Now add pseudoenergies so that this 
becomes a symmetry of the 
whole, and then check that this enlarged system does indeed
give the desired value, $2(N{-}1)/(N{+}2)$, for the ultraviolet 
central charge. For $N$ even the magnonic structure is rather trivial
and the 
results are already contained in equations~(\ref{Eziv}),(\ref{TBAziv})
and (\ref{ENzbn}),(\ref{TBAzbn}). However the idea also works for every odd
$N$; the resulting systems will now be given.

When $N{=}4n{+}1$, the magnonic structure of the sine-Gordon TBA is
identical to that at $N{=}5$, as given in equation~(\ref{TBAzzV}) -- the
only differences come from the appearance of $n{-}1$ breathers in the
spectrum in addition to the fundamental soliton-antisoliton doublet.
This
leads to the following proposal for an $N{=}4n{+}1$ TBA system:
\eq
RE(R)=-{1\over 2\pi}\!\sum^{{}\atop n}_{i=1\atop\alpha=1,6}
 \iint\!d\theta\,\nu_i^{(\alpha)}(\theta)L^{(1)}_i(\theta)
\label{ENzvgen}
\en
with
\bea\ds
\ep^{(\alpha)}_i(\theta)&=&\nu^{(\alpha)}_i(\theta)-
\!\sum^n_{j=1} \lf[\phi_{ij}{*}L^{(\alpha)}_j(\theta)
-\psi_{ij}{*}L^{(\wtilde\alpha)}_j(\theta)\ri]\nn\\
&&~~~{}-\delta_{i,n}\!\sum^6_{\beta=1}l_{\alpha\beta}^{[e_6]}
           \phi_2{*}L^{(\beta)}_n(\theta)\, ,~~~~\,(\alpha{=}1,6,~
            \wtilde\alpha=7{-}\alpha)\,;\nn\\
{}~\ep^{(\alpha)}_n(\theta)&=&{~~~}
\sum^6_{\beta=1}l^{[e_6]}_{\alpha\beta}
  \phi_2{*}L^{(\beta)}_n(\theta)\, ,~~~~~~~~~\quad(\alpha=2\dots 5)\,.
\label{TBAzvgen}\\
\nn
\eea
\vskip -5pt
\noindent
The energy terms are $\nu_i^{\sscr(1)}(\theta)=\nu_i^{\sscr(\tilde
1)}(-\theta)=\half M_iRe^{\theta}$, with the $M_i$ given by
eqs.~(\ref{bmasses}),(\ref{bbmasses}) with $h{=}N/2{-}2$, and
the kernels are given 
by equations~(\ref{kernels})--(\ref{magkernels}).

Finally, for $N{=}4n{+}3$ the sine-Gordon magnonic structure mimics that 
for $N{=}7$, $\beta^2{=}32\pi/7$. The system
can therefore be found simply by tacking the
appropriate number of breather-like pseudoenergies onto 
equations~(\ref{TBAzvii}), (\ref{ENzvii}). With energy
terms and kernels extracted from appendix~A as before, it reads
\eq
RE(R)=-{1\over 2\pi}\!\sum^{{}\atop n}_{i=1\atop\alpha=1,6}
 \iint\!d\theta\,\nu_i^{(\alpha)}(\theta)L^{(\alpha)}_i(\theta)
\label{ENzviigen}
\en
\vskip -5pt\noindent
with
\bea\ds
\ep^{(1)}_i(\theta)&=&\nu^{(1)}_i(\theta)-
\!\sum^n_{j=1} \lf[\phi_{ij}{*}L^{(1)}_j(\theta)
-\psi_{ij}{*}L^{(6)}_j(\theta)\ri]\nn\\
&&~~~{}-\delta_{i,n}\lf[\phi_3{*}(L^{(4)}_n(\theta)+L^{(5)}_n(\theta))
+\phi_4{*}L^{(2)}_n(\theta)+\phi_5{*}L^{(3)}_n(\theta)\ri]\nn\\
\ep^{(2)}_n(\theta)&=&~~\phi_2{*}(K^{(3)}_n(\theta)-L^{(1)}_n(\theta))\nn\\
\ep^{(3)}_n(\theta)&=&~~
  \phi_2{*}(K^{(2)}_n(\theta)+K^{(4)}_n(\theta)+K^{(5)}_n(\theta))~~;~~~
 \ep^{(4)}_n(\theta)=\phi_2{*}K^{(3)}_n(\theta)\nn\\
\ep^{(5)}_n(\theta)&=&~~\phi_2{*}(K^{(3)}_n(\theta)-L^{(6)}_n(\theta))\nn\\
\ep^{(6)}_i(\theta)&=&\nu^{(6)}_i(\theta)-
\!\sum^n_{j=1} \lf[\phi_{ij}{*}L^{(6)}_j(\theta)
-\psi_{ij}{*}L^{(1)}_j(\theta)\ri]\nn\\
&&~~~{}-\delta_{i,n}\lf[\phi_3{*}(L^{(4)}_n(\theta)+L^{(2)}_n(\theta))
+\phi_4{*}L^{(5)}_n(\theta)+\phi_5{*}L^{(3)}_n(\theta)\ri]~.\nn\\
\label{TBAzviigen}
\eea

\subsection{Modifications for massive flows}
In their ultraviolet limits, $\MR\ll 1$, all of the TBA equations
again simplify, though in a rather different manner to the infrared
situation already described. In a central region $-\log(1/\MR)\ll
\theta\ll\log(1/\MR)$ the energy terms $\nu_i^{\sscr(\alpha)}(\theta)$ 
can be ignored and the pseudoenergies are
approximately constant. Then, in the neighbourhoods of
$\theta{=}{\pm\log(1/\MR)}$, the equations
reduce to kink systems of a new type, different from those encountered
in the infrared in that they continue to involve {\it all} of the
pseudoenergies. They are obtained from the original TBA
equations by neglecting either the energy terms proportional to
$\MR e^{\theta}$ (this near $\theta{=}{-\log(1/\MR)}$), or else
the terms proportional to $\MR e^{-\theta}$ (this
near $\theta{=}{+\log(1/\MR)}$). If the solutions to these two kink
systems are denoted $\epL_i{}^{(\alpha)}$ and 
$\epR_i{}^{(\alpha)}$ respectively, then as $R\rightarrow 0$
\eq
\opnup{1}
\ep^{(\alpha)}_i(\theta)\sim \left\{ \begin{array}{ll}
\epL_i{}^{(\alpha)}(\theta)&{}~~(\theta\approx -\log{1\over \MR}~)\\[1pt]
\epR_i{}^{(\alpha)}(\theta)&{}~~(\theta\approx +\log{1\over \MR}~) 
\end{array} \right.
\label{kinks}
\en 
Once these two kink systems have
decoupled, all scale dependence is lost from the equations determining
the effective central charge, and with the aid of the dilogarithm
identities listed in appendix~C, the value of $c(0)$ can be extracted. The
result is $2(N{-}1)/(N{+}2)$; and this, together with the result
$c(\infty){=}1$ already found, is exactly
as would be expected of a flow from the
Fateev-Zamolodchikov multicritical point into the Kosterlitz-Thouless
phase. Subsequent sections will give further evidence for this
interpretation.  But first, the picture should be completed with a set 
of ans\"atze for the opposite perturbation, into the massive phase.
There is a standard procedure to go from a massless to a 
massive TBA 
system: simply remove all energy terms $\nu^{\sscr(\wtilde 1)}_i$
proportional to $\halfs\MR e^{-\theta}$, and in the remaining terms 
$\nu_i^{\sscr(1)}$ replace $\halfs\MR e^{\theta}$ with $\MR\cosh\theta$ 
throughout. Up to a relabelling of the
nodes $\alpha{\rightarrow}\wtilde\alpha$ for the kink system near
$\theta{=}{-\log(1/\MR)}$, the leading ultraviolet behaviour of the
equations is unchanged by this man\oe uvre, and so the result $c(0)
=2(N{-}1)/(N{+}2)$ is preserved. However,
the decoupling previously induced
by the energy terms in the large $R$ limit no longer occurs, and
instead one finds $c(\infty)=0$, as required for a massive flow. 
These systems are thus good 
candidates to describe the situation for the opposite sign of the
coupling constant, and their properties will also be investigated
below. It is interesting that after this final step, the sine-Gordon
scaffolding, by which we levered ourselves up from $c{=}1$ to
$c{=}2(N{-}1)/(N{+}2)$, has been completely removed.

\bigskip

\noindent
Without a prior knowledge of the $\ZZ_N$ conformal field
theories, the notion to
reinterpret suitable sine-Gordon TBA systems as the infrared limiting
forms of larger sets of equations  might conceivably have led to 
their rediscovery. It is therefore worth
investigating whether there are any other values of $\beta^2$ at
which this might work. 
Curiously, it seems that the points $\beta^2=32\pi/N$
already captured really are special in this regard -- for no other
values of the coupling constant does the magnonic part of the sine-Gordon
TBA system exhibit a $\ZZ_2$ symmetry with the desired properties. The only
other option is to aim for an enlarged system in which the left-right
$\ZZ_2$ symmetry acts trivially on the magnonic pseudoenergies. Formally,
this
can be done at {\it any} value of $\beta^2$: in the repulsive regime
($\beta^2>4\pi$; no breathers) it leads to sausage models~\cite{FOZa}
in every case. 
In the attractive regime the picture is not so straightforward. For
example, the most
natural choice of kernels to link the doubled-up nodes,
namely the functions $\psi_{ij}$ successfully used above, does not
seem  to yield any sensible (that is, rational) values for the ultraviolet 
central charges. This is probably related to the `intolerable' properties
that the massless sausage S-matrices
acquire as the attractive regime is entered~\cite{FOZa}.

\resection{Ultraviolet asymptotics}

If the functions $E(R)$ defined in the last section really
are the ground-state energies of certain perturbed conformal field
theories, then their behaviours at small $R$ should be consistent with
the results of conformal perturbation theory. For a unitary 
theory perturbed by a primary operator $\epsilon$ of dimensions
$(\Delta,\Delta)$, this predicts the expansion
\eq
E^{\rm (pert)}(\lambda,R)=-{\pi c\over 6R}
 +{2\pi\over R}\sum_{m=2}^{\infty}B_mt^m~.
\label{Epert}
\en
The coupling $\lambda$ appears in the dimensionless
parameter
$t=-2\pi\lambda \lf({R/2\pi}\ri)^{2-2\Delta}$, while the
conveniently-normalised coefficients $B_m$ are given in terms of the
connected correlation functions  of the perturbing field $\epsilon$ on
the plane:
\eq
B_m=-{1\over (2\pi)^{m{-}1}m!}\int
\VEV{\epsilon(1,1)\epsilon(z_1,\bar z_1)\dots\epsilon(z_{m-1},\bar
z_{m-1})}_C\prod^{m-1}_{k=1}{d^2z_k\over (z_k\bar z_k)^{1{-}\Delta}}~.
\label{Bndef}
\en
The conformal mapping $z=e^{-2\pi w/R}$, from the $w$-cylinder of
circumference $R$ to the infinite $z$-plane, is responsible for the 
appearance of the conformal anomaly $c$ in~(\ref{Epert}) and also for the
measure used in~(\ref{Bndef}); see for example
ref.~\cite{Zb}. If $2\Delta\geq 1$, then a finite number of terms,
up to order $n\leq 1/(1{-}\Delta)$, suffer from
ultraviolet divergences which must be regularised: the usual
prescription is to analytically continue in $\Delta$ from a region
where the relevant integrals converge. Sometimes even this fails to
render the answer finite, a sign of an irregular term in the
expansion of~$E^{\rm (pert)}$.

The constraints of conformal invariance are such that $B_2$ and $B_3$
are completely fixed by the values of $\Delta$ and the operator
product coefficient $C_{\epsilon\epsilon\epsilon}$, irrespective of
any other details of the theory. For later reference, they
are~\cite{DFa}
\eq
B_2=-{1\over 4}{\gamma(\Delta)^2\over\gamma(2\Delta)}\quad,\quad
B_3=-{C_{\epsilon\epsilon\epsilon}\over 48}
{\gamma(\Delta/2)^3\over\gamma(3\Delta/2)}~,
\label{bform}
\en
where $\gamma(x)=\Gamma(x)/\Gamma(1{-}x)$. 
For the particular models under discussion, the values of
$C_{\epsilon\epsilon\epsilon}$ were found by Zamolodchikov and
Fateev to be~\cite{ZFa}:
\eq
C_{\epsilon\epsilon\epsilon}^{\sscr(N)} = \frac{2}{35}
\frac{\Pi^{\sscr(N)}(7) \Pi^{\sscr(N)}(2)^{3}}{\Pi^{\sscr(N)}(4)}
\sqrt{5^{3} \frac{\Gamma(\frac{N+3}{N+2})\Gamma(1-\frac{5}
{N+2})^{3}}{\Gamma(\frac{N+1}{N+2})  \Gamma(1+\frac{5}{N+2})^{3} }  }
\label{ceee}
\en
where $\Pi^{\sscr(N)}(j)=\prod_{k=1}^{j} \frac{\Gamma(1+\frac{k}{N+2})}
{\Gamma(1-\frac{k}{N+2})}$.

The perturbative expansion is expected to have a non-zero radius of
convergence, and the function that it defines should behave at large $R$
as $E^{\rm (pert)}(\lambda,R)\sim \Epsilon(\lambda)R$, 
$\Epsilon(\lambda)$ being 
the bulk contribution to the ground-state energy. Whilst this is 
subtracted off in the TBA, 
this is not the case in conformal perturbation theory
and  $\Epsilon(\lambda)$ may be nonzero.
To take this into account, the
predictions of the last section should be compared not with $E^{\rm
(pert)}$ directly, but rather with the subtracted quantity
\eq
E(\lambda,R)=E^{\rm (pert)}(\lambda,R)-\Epsilon(\lambda)R~.
\label{ETBA}
\en
Finally, note that the functions $E(R)$ of the
last section all in fact depend on $R$ and $M$, where $M$ is a
mass scale, part of the infrared specification of the model. (In the 
massive direction, $M$ will turn out to be
the mass of a certain asymptotic one-particle state; in the massless
direction it sets the crossover scale.) Conformal perturbation theory
sees instead $R$ and the coupling constant $\lambda$. However, on 
dimensional grounds there must be a relation between $\lambda$ and $M$ 
of the form
\eq
\lambda=\kappa M^y~,
\label{kappadef}
\en
where $y=2{-}2\Delta$ is the dimension of $\lambda$, and $\kappa$ is
some dimensionless constant relating the short and long distance
descriptions of the theory. Later we will conjecture an
exact expression for $\kappa$, but for the moment it must be
considered as another unknown.

Returning now to the equations of the last section, the first piece of
analysis, following Zamolodchikov~\cite{Zf}, determines the value 
of $y$. This relies on the fact that any solution to one of the TBA
systems above automatically furnishes a solution 
\[
Y_i^{(\alpha)}(\theta)\equiv e^{\ep_i^{(\alpha)}(\theta)}
\]
to a set of functional relations known as a Y-system. The
Y-systems appropriate to the problem in hand are given in
appendix~B, and numerically they turn out to entail the following 
periodicity property for their solutions:
\eq
Y_i^{(\alpha)}\lf(\theta+2\pi\imath{N{+}2\over N{-}4}\ri)
= Y_i^{(\alpha)}\lf(\theta\ri)~.
\label{yfulper}
\en
As a result, the $Y_i^{(\alpha)}$ have an expansion in powers of
$u(\theta)=e^{{N{-}4\over N{+}2}\theta}$. For the full TBA
system this will be in positive and negative powers, since in
the full system energy terms blow up both at $\theta{=}{-\infty}$
($u{=}0$), and at $\theta{=}{+\infty}$ ($u{=}\infty$). 
However for the kink systems which govern
the ultraviolet limit, energy terms diverge in one direction only
and so $\YR_i{}^{(\alpha)}=\exp{\epR{}_i^{(\alpha)}}$ is regular at $u{=}0$,
and $\YL_i{}^{(\alpha)}$ is regular at $u{=}\infty$. Therefore 
$\epR_i{}^{(\alpha)}$ has an expansion in 
non-negative powers of $u$ about $u{=}0$, 
and $\epL_i{}^{(\alpha)}$ an expansion in non-positive powers about
$u{=}\infty$. This information is
important, because a part of the corrections to the asymptotic form of 
each kink
solution can be traced to the presence of the other kink at a finite
distance in $\theta$, namely $2\log(1/\MR)$. If this is incorporated in
an iterative fashion, then the corrections will enter as
powers of $u(2\log(1/\MR))$. In turn, this implies that in the TBA
corrections to $RE(R)$ form a regular series:
\eq
RE(R)=-{\pi c\over 6}
     +2\pi\sum_{m=1}^{\infty}F_m\,(\MR)^{2{N{-}4\over N{+}2}m}~,
\label{regexp}
\en
modulo further terms, to be mentioned shortly, that can be traced
to the subtraction of the bulk term performed in 
equation~(\ref{ETBA}).
This agrees with the result~(\ref{Epert}) from conformal
perturbation theory so long as $y=2(N{-}4)/(N{+}2)$. This gives 
a first check on the TBA: the prediction
\[
\Delta={6\over N+2}
\]
for the conformal dimension of the perturbing field. From
the work of Zamolodchikov and Fateev~\cite{ZFa}, there is indeed 
a suitable scalar operator in the $\ZN$-symmetric conformal theory:
sometimes denoted $\epsilon^{(2)}$, it is both
$\ZN$-neutral and self-dual -- precisely the two properties that were
demanded in the later part of section~1.
Furthermore, Fateev~\cite{Fa} has shown that this perturbation is
integrable, so the existence of a TBA system describing it is to be
expected.

Next we turn to the bulk piece, which, via equation~(\ref{ETBA}), should
show 
itself in the ultraviolet as a term $\Epsilon(\lambda)R^2$ in the
expansion of $RE(R)$.
Following refs.~\cite{Zb,KMc}, it is possible
to extract the exact value of the coefficient $\Epsilon$, as a
function of $M$, from the TBA.
The result, which depends on whether the system is massive or
massless, is
\eq
\Epsilon_{\rm massive}=
 -{M^2\over 2}{\sin\theta_N\sin 2\theta_N \over\sin 3\theta_N}
\quad,\quad
\Epsilon_{\rm massless}=
 -{M^2\over 2}{\sin^2\theta_N\over\sin 3\theta_N}~,
\label{bulkterms}
\en
where $\theta_N=2\pi/(N{-}4)$. This formula can be derived generally
for $N{\ge}8$, and was also checked by specific calculations for
$N{=}5$ and $N{=}6$.
When $N{=}4n{+}4$, the value of $\Epsilon_{\rm massive}$ agrees with
an earlier result~\cite{Fa} for the massive $\phi_{\rm 1,1,Adj}$ 
perturbation of the 
$d_{n{+}1}^{(2)}{\times}d_{n{+}1}^{(1)}/d_{n{+}1}^{(3)}$ coset model;
this follows the correspondence~(\ref{cosets}) already
mentioned for some of the massless flows.

There are two values of $N$ for which these formulae break
down: if $N{=}7$ or $10$, the results are formally infinite.
This infinity should
cancel against a term in the regular expansion~(\ref{Epert}),
to render the overall result finite~\cite{HFa}.
For the $R$-dependencies to match
up, $m$, the order of the putative cancelling term, must be equal to 
$(N{+}2)/(N{-}4)$; this is indeed an integer for $N{=}7$ and $10$.
(Such a resonance is also found in the $N{=}5$
theory, between the bulk term and the term of order $\lambda^{7}$ in the
regular expansion, but~(\ref{bulkterms}) shows that the divergence is
anyway absent in this case.) The dominant residual piece can be
extracted by evaluating (in the massive case)
\[
\lim_{N\rightarrow2{2m{+}1\over m{-}1}}\lf[
 -{\sin\theta_N\sin 2\theta_N \over 2\sin 3\theta_N}
\lf((\MR)^2-(\MR)^{2{N{-}4\over N{+}2}m}\ri)\ri]
\]
and then setting $m=(N{+}2)/(N{-}4)$ at the end. This gives
\eq
\Epsilon_{\rm massive}=
 -M^2{\sin\theta_N\sin 2\theta_N \over\pi\cos 3\theta_N}
{N{-}4\over N{+}2}\log\MR~.
\label{genlogbulkterms}
\en
The calculation for the massless case is almost
identical: just replace $\sin
2\theta_N$ by $\sin\theta_N$ throughout. The two instances of these
formulae pertinent to the current discussion give us
\bea
\Epsilon_{\rm massive}^{(N{=}7)}=\phantom{-}
{1\over 4\pi}M^2\log\MR\quad &,&\quad
\Epsilon_{\rm massive}^{(N{=}10)}={3\over 8\pi}M^2\log\MR\nn\\
\Epsilon_{\rm massless}^{(N{=}7)}=-{1\over 4\pi}M^2\log\MR\quad &,&\quad
\Epsilon_{\rm massless}^{(N{=}10)}={3\over 8\pi}M^2\log\MR\nn\\
\label{logbulkterms}
\eea
It is also possible to say something about the nonsingular parts of
the bulk terms, which have so far been disregarded: see
equation~(\ref{finbulkterms}) below.

These results have been obtained by analytic continuation in $N$, back
to the `critical' values of 7 and 10. But it should also be possible
to see the logarithms directly in conformal
perturbation theory, at order
$\lambda^3$ when $N{=}7$ and $\lambda^2$ when $N{=}10$. Note how this
is already reflected in the relative
signs of the massive and massless results in~(\ref{logbulkterms}). To
say more, the integrals~(\ref{Bndef}) must be re-examined, since
their values are needed at precisely the points where the general
formulae~(\ref{bform}) break down. The relevant integrals acquire
logarithmic divergences at short distances, though not in fact at long
distances: the `metric factors' $1/(z_k\bar z_k)^{1{-}\Delta}$ prevent
this. To cope with the short-distance
divergences, an explicit cutoff
$a$ can be introduced, with the understanding that when the
perturbative integrals are done on the cylinder, no two operators
should get any closer than $a$. Transforming this onto the
plane, via $z=e^{2\pi w/R}$, to leading order in $a$ operators are now
required to stay  at least $2\pi a/R$ apart. This can be implemented
by inserting Heaviside step functions
$\theta_{|x|}\equiv\theta(|x|{-}{2\pi a\over R})$ as appropriate.
Since the interest is in the 
limit where all the $z_k{\rightarrow}1$, the metric factors
can be ignored and replaced with an
explicit cutoff in the infrared. Then 
the logarithmic divergence in $B_2$, which occurs 
when $N{=}10$ and $\Delta{=}1/2$, is found by
\bea
\opnup{1}
B_2&=&-{1\over 4\pi}\int^{|z|<1}
\VEV{\epsilon(0,0)\epsilon(z,\bar z)}_C\,\theta_{|z|}\,d^2z+{\rm n.s.}
\nn\\[1pt]
&=&-{1\over 2}\int_{2\pi a/R}^1 r^{-1} dr +{\rm n.s.}\nn\\[1pt]
&=&-{1\over 2}\log{R\over 2\pi a}+{\rm n.s.}\label{btwo}\\ \nn
\eea
\vskip -5pt\noindent
Here, `n.s.' denotes the non-singular terms which do not contribute to
the logarithm. Similarly, when $N{=}7$, $\Delta{=}2/3$ and there is a 
logarithmic divergence in $B_3$. This can be extracted as follows:
\bea
\opnup{1}
B_3&=&-{1\over 24\pi^2}\int^{|x|<1\atop |y|<1}\!
\VEV{\epsilon(0,0)\epsilon(x,\bar x)\epsilon(y,\bar y)}_C\,
\theta_{|x|}\theta_{|y|}\theta_{|x{-}y|}d^2x\,d^2y +{\rm n.s.}\nn\\[2pt]
&=&-{C^{\sscr(N{=}7)}_{\epsilon\epsilon\epsilon}
\over 24\pi^2}\int^{|x|<1\atop |y|<1}\!\!
{\theta_{|x|}\,\theta_{|y|}\,\theta_{|x{-}y|}\,d^2x\,d^2y \over
[x\bar x y\bar y(x{-}y)(\bar x{-}\bar y)]^{2/3}} +{\rm n.s.}\nn\\[2pt]
&=&-{C^{\sscr(N{=}7)}_{\epsilon\epsilon\epsilon}
\over 24\pi^2}\int^{\sscr |x|<1}\!\!{\theta_{|x|}\over [x\bar x]}
\lf[\int^{\sscr |y|<1/|x|}\!\!\!{\theta_{|xy|}\,\theta_{|x-xy|}\,d^2y
\over [y\bar y(1-y)(1-\bar y)]^{2/3}}\ri]d^2x+{\rm n.s.}\nn\\ \nn
\eea
\vskip -5pt\noindent
Now the integral in square brackets, call it $I(x,a)$, is convergent
(and non-zero) at $x{=}a{=}0$, and so to leading order it can be
replaced by $I(0,0)$, which is equal
to $\pi\gam{{1\over 3}}^3$~\cite{DFa}. The
remaining integral gives the logarithm, just as in
equation~(\ref{btwo}). Collecting the pieces together,
\eq
B_3=-{1\over 12}\, C^{\sscr(N{=}7)}_{\epsilon\epsilon\epsilon}
\gam{{1\over 3}}^3\log{R\over 2\pi a}+{\rm n.s.}~,
\label{bthree}
\en
where $C^{\sscr(N{=}7)}_{\epsilon\epsilon\epsilon}$ 
is given by equation~(\ref{ceee}).

Since these calculations have left the nonsingular parts of $B_2$ and
$B_3$ completely uncontrolled, it would appear that no prediction
can be made about the contribution to $RE(R)$ that is proportional to 
$R^2$ itself. However this is not quite true: assuming that the same
regularisation scheme
is used in the massless and massive directions, all of the
perturbative contributions to $\Epsilon^{(N)}_{\rm massless}$ and 
$\Epsilon^{(N)}_{\rm massive}$  must cancel out when their sum is
taken for $N{=}7$, or their difference for $N{=}10$. The residual
finite pieces then come entirely from~(\ref{bulkterms}), and are:
\eq
\Epsilon_{\rm massive}\pm
\Epsilon_{\rm massless}=\pm{\sqrt{3}\over 8} M^2~.
\label{finbulkterms}
\en
The plus signs apply when $N{=}7$; the minus signs when $N{=}10$.

Returning to the singular terms, substituting
equations~(\ref{btwo}) and (\ref{bthree}) into the general
expansion~(\ref{Epert}) gives new expressions for the
logarithmically-divergent pieces of the bulk terms. But whereas the
earlier expressions, equations~(\ref{logbulkterms}), were in terms of
$M$, this time the coupling constant $\lambda$ appears.
Equating the alternatives allows the constant $\kappa$ appearing in the
relation~(\ref{kappadef}) to be determined exactly, at least for these two
values of $N$. The results turn out not to depend on whether the
perturbations are in the 
massive or the massless directions, and are:
\bea
N{=}7\,:\quad\kappa^3&=&{3\over
4 \pi^3\gam{{1\over 3}}^3
C^{\sscr(N{=}7)}_{\epsilon\epsilon\epsilon} }~~;\nn\\
N{=}10\,:\quad\kappa^2&=&{3\over 8\pi^2}~~.
\label{blobb} \\ \nn
\eea

\vskip -7pt\noindent
There are other values of $N$ for which $\kappa$ is known exactly: a
special case of a formula due to Fateev gives its value for the
massive flow, whenever
$N{=}4n$~\cite{Fb}. It turns out that when continued to the
fractional values $n{=}7/4$ and $n{=}5/2$, the results~(\ref{blobb})
are reproduced. This motivates the conjecture that the analytic
continuation of Fateev's result holds for {\it all} values of $N$, and
for both the massive and the massless perturbations:
\eq
{\rm Any~}N\,:\quad
\kappa^2={4\over 9\pi^2}\,{\gam{{4\over N{+}2}}^2\gam{{5\over
N{+}2}}\over\gam{{1\over N{+}2}}}\lf[{\pi\Gam{{N{+}2\over N{-}4}}\over
\Gam{{2\over N{-}4}}\Gam{{N\over N{-}4}}}\ri]^{4{N{-}4\over N{+}2}}.
\label{blobgen}
\en
It is not meant to be obvious that the formulae (\ref{blobb}) are
special cases of this! Further support for the conjecture comes
from numerical results, the subject of the remainder of this section.

One reason for undertaking a numerical study of the TBA equations 
is the lack of any analytic results for the coefficients $F_m$
in the expansions (\ref{regexp}). In particular, to be sure that the
massive and massless TBA systems really do describe perturbed
conformal theories differing only in the sign of a coupling
constant $\lambda$, the relative signs of the $F_m$'s predicted
for each pair
of flows should be examined. Letting $F_m$ denote the coefficients
for the massless flow, and $\wtilde F_m$ those for the massive
flow, and accepting provisionally the conjecture that the mass scales 
are equal in the massive and massless directions,
the expectation is 
\eq
F_{m}=(-1)^m\wtilde F_{m}~.
\label{expexp}
\en

An iterative method was used to solve the TBA equations for 
$N{=}5$, $6$, $7$, $9$ and $10$, discretising the $\theta$-axis in steps 
of $\delta\theta{=}0.1\,$ and normalising $M{=}1$. 
Direct iteration fails to converge~\cite{KMb}; adding
$\ep^{(\alpha)}_i$ to both sides of each
equation, dividing by two and basing the iteration on
the result solves this problem, though
convergence is rather slow (up to 300 iterations 
were used at each value of $R$). Results for $RE(R)$
to about 14 digits precision  were obtained
at $30$ values of $(\MR)^y$ between $0.03$ and $0.5$. 
The exact constant piece, namely $-\pi (N{-}1)/3(N{+}2)$, was then
subtracted, 
along with the bulk terms as predicted by equations~(\ref{bulkterms})
and~(\ref{logbulkterms}).
In addition, for $N{=}7$ and $10$, 
the unambiguously nonperturbative contributions to the bulk terms, 
identified using equation~(\ref{finbulkterms}), were subtracted in a
symmetrical way. 
The remainder of $RE(R)$ after all this should in each case
have a regular expansion in $(\MR)^y$, with coefficients that conform
to the general rule~(\ref{expexp}). A fit against the numerical data 
was consistent with this, and the values of the first few
$F_m$ and $\wtilde F_m$  thereby obtained are reported in
appendix~D. Note that the TBA system for $N{=}8$ is simply a doubling-up
of Zamoldchikov's proposal for the tricritical Ising to Ising
flow~\cite{Zd}. The expansion coefficients for $N{=}8$ are therefore
easily obtained from the earlier data, and are also included.
In our fits, the $m{=}0$ and $m{=}1$ coefficients were left 
free; comparing their measured values with the exact predictions that
they be zero gives a check on the numerical accuracy.

Besides the clear agreement of the results with 
equation~(\ref{expexp}), a number of other features deserve mention.

First, to within the numerical errors the $\ZZ_5$ TBA has $\wtilde F_3=
F_3=0$. This matches with the perturbative expansion: when $N{=}5$,
$\Pi^{\sscr(N)}(7)=0$ and, from equation~(\ref{ceee}), 
$C_{\epsilon\epsilon\epsilon}^{\sscr(5)}$ vanishes.

\begin{table}[thb]
\widetable
\begin{center}
\begin{tabular}{|c|l|l|l|}     \hline \hline
\rll
{}~$N$~{}&$~~~~B_2^3/B_3^2$&$~~~~\wtilde F_2^3/\wtilde F_3^2$&
$~~~~F_2^3/F_3^2$ \\
\hline
6 &~\, 0.046260423~~ &~\, 0.046260427~~ &~\, 0.046260424~~ \\
\hline
8 &~\, 0.379827746 &~\, 0.379827746 &~\, 0.379827746\\
\hline
9 &~ 14.448969 &~ 14.448967 &~ 14.448991\\
\hline \hline
\end{tabular}
\end{center}
\vskip -10pt
\caption{Comparison of the $\ZZ_6$, $\ZZ_8$ and $\ZZ_9$ systems with 
conformal perturbation theory}
\label{z6comp}
\end{table}
%

For $N{=}6$, $8$ and $9$ a more delicate test can be done. 
Independently of the
value of $\kappa$, $F_2^3/F_3^2=\wtilde F_2^3/\wtilde F_3^2=
B_2^3/B_3^2$ should hold if the TBA is to be consistent with
conformal perturbation theory. Using equations~(\ref{bform})
and~(\ref{ceee}) to evaluate $B_2^3/B_3^2$, the numerical results are
indeed consistent with this, as table~\ref{z6comp} demonstrates.

Finally, the results for $N{=}7$ and $10$ support the
formulae~(\ref{logbulkterms}) and (\ref{finbulkterms}): any errors in
the bulk terms subtracted would have destroyed the good agreement
that the fits show with the prediction~(\ref{expexp}).

\begin{table}[htb]
\widetable
\begin{center}
\begin{tabular}{|c|c|c|c|}     \hline \hline
\rll
{}~$N$~{} & $\kappa^2_{\rm exact}$ &
$\wtilde\kappa^2_{\rm num}$ & $\kappa^2_{\rm num}$ \\
\hline
5&~0.0235204664 & 0.0235204667 & 0.0235204663 \\
\hline
6&0.0371477546 & 0.0371477547 & 0.0371477547 \\
\hline
7&0.0434800500 & 0.0434800501 & 0.0434800501 \\
\hline
8&0.0442207130 & 0.0442207131 & 0.0442207131 \\
\hline
9&0.0418094000 & 0.0418094001 & 0.0418094003 \\
\hline
10&0.0379954439 & 0.0379954451 & 0.0379954487 \\
\hline \hline
\end{tabular}
\end{center}
\vskip -10pt
\caption{Comparison of exact and numerical values for $\kappa^2$}
\label{kappacomp}
\end{table}

Once the consistency of the data with conformal perturbation theory
has been established, numerical estimates for $\kappa$ can be extracted 
and checked against the conjecture~(\ref{blobgen}). Equating the
terms proportional to $R^{2y}$ in equations~(\ref{Epert})
and~(\ref{regexp}) yields
\[
\kappa^2_{\rm num}=(2\pi)^{2y-2}{F_2\over B_2} =
-{(2\pi)^{2y}\over\pi^2}{\gam{{12\over N{+}2}}\over
\gam{{6\over N{+}2}}}F_2~,
\]
and similarly for $\wtilde\kappa^2_{\rm num}$. For $N{=}10$, the
second-order ($m{=}2$) term is obscured by the
logarithm, but a prediction for $\kappa$ can be
obtained in an analogous way from the third-order behaviour.
The numbers obtained in this way are compared with the exact 
predictions, $\kappa^2_{\rm exact}$, in table~\ref{kappacomp}; the
agreement is very good.

%
\resection{Infrared limits}
 
The behaviours of the massless and massive systems become very
different as $R$ grows, and they will be described
separately. We start with the massless flows.

\subsection{Into the Kosterlitz-Thouless phase}

The massless systems all exhibit $c(\infty){=}1$, a prerequisite if
the picture outlined at the end of section~1 is to be confirmed. For
a more detailed check, some information on the possible operator
contents of each $\ZZ_N$ model in its Kosterlitz-Thouless phase is
needed. A complete description can be found in ref.~\cite{vGRSa}:
the scaling limit of the model is
equivalent to a free boson compactified on a circle of some radius
$r$. Adopting the normalisation traditional in conformal field
theory, the action can be taken as
\eq
S^*_{\rm IR}[\Phi]=\frac{1}{2\pi}\int d^2x (\partial_{\mu}\Phi)^2\,.
\label{bosact}
\en
The boson $\Phi$ can be split into holomorphic and antiholomorphic 
components by writing $z=x_1{+}\imath x_2\,$, 
$\Phi(z,\bar z)=\half(\phi(z){+}\bar\phi(\bar z))\,$, with
\[
\VEV{\phi(z)\phi(w)}=-\log(z{-}w)~~,~~
\VEV{\bar\phi(\bar z)\bar\phi(\bar w)}=-\log(\bar z{-}\bar w)~.
\]
Compactifying onto a circle of radius $r$ amounts to the
identification $\Phi\equiv\Phi{+}2\pi r$. Closer to the discussion of
section~1 would have been to work with $\theta(z,\bar z)\equiv
\Phi(z,\bar z)/r$,
whereupon the prefactor in~(\ref{bosact}) would have become
$r^2/2\pi$. This shows that $r^2/\pi$ should be thought of as the 
inverse temperature.

Invariance under $\Phi\rightarrow\Phi{+}2\pi r$ and mutual locality
restrict the possible vertex operator primary fields to 
\[
V_{nm}^{+}(z,\bar z)
 =\sqrt{2}\cos(p\phi(z){+}\bar p\bar\phi(\bar z))~~,~~~
V_{nm}^{-}(z,\bar z)
 =\sqrt{2}\sin(p\phi(z){+}\bar p\bar\phi(\bar z))~,~~~
\]
where $(p,\bar p)=(\frac{n}{2r}{+}mr,\frac{n}{2r}{-}mr)$ and $n,m\in\ZZ$.
These have the conformal weights
\[
(\Delta_{nm},\bar\Delta_{nm})=(\halfs p^2,\halfs\bar p^2)
=(\halfs(\frac{n}{2r}{+}mr)^2,\halfs(\frac{n}{2r}{-}mr)^2)~.
\]
Included in this collection are the the vortex fields $V^{\pm}_{0M}$,
and the $N$-fold symmetry breaking fields
$V^{\pm}_{N0}$.  Their conformal weights are:
\[
\Delta_{N0}(r)=\bar\Delta_{N0}(r)=\frac{N^2}{8r^2}\quad,\quad
\Delta_{0M}(r)=\bar\Delta_{0M}(r)=\frac{M^2r^2}{2}~.
\]
This information allows the of values of $r$ actually realised
in the Kosterlitz-Thouless region to be determined.
At the low-temperature (large-$r$) border, the fields
$V^{\pm}_{N0}$ become relevant, so $\Delta_{N0}(r_{\rm max})=1$, 
and $r_{\rm max}=N/\sqrt{8}$. Similarly the high-temperature
boundary is encountered when the least irrelevant vortex fields,
$V^{\pm}_{01}$, become relevant, that is when $\Delta_{01}(r)=1$, so
$r_{\rm min}=\sqrt{2}$. One further value of $r$ can be identified 
in the thermodynamic phase space. The duality transformation for the
$\ZZ_N$ model exchanges the $N$-fold symmetry-breaking field with the
unit vortex, that is $V_{N0}$ with $V_{01}$. 
Thus $r_{\rm sd}$, the compactification radius appropriate to any
self-dual system, must satisfy $\Delta_{N0}(r_{\rm
sd})=\Delta_{01}(r_{\rm sd})$. Hence, $r_{\rm sd}=\sqrt{N/2}$.
To summarise:
\vskip -4pt
{
\renewcommand{\arraystretch}{1.2}
\begin{center}
\begin{tabular}{c|cc}
$r$ &{}~~ $\Delta_{N0}$ & $\Delta_{01}$ \\
\hline
$r_{\rm min}{=}\sqrt{2}$&$\fracs{N^2}{16}$&$1$ \\
$r_{\rm sd}{=}\sqrt{\fracs{N}{2}}$&$\fracs{N}{4}$&$\fracs{N}{4}$ \\
$r_{\rm max}{=}\fracs{N}{\sqrt{8}}$&$1$&$\fracs{N^2}{16}$ \\
\end{tabular}
\end{center}
}
\noindent
Since initial (ultraviolet) perturbation was self-dual, it is the
middle row of the table that will be important.
The discussion has been couched in the language of 
continuum-valued spins, and it might appear surprising that a model with
discrete spins can mimic such behaviour. Universality and
a suitable renormalisation group should anyway suffice, but for
a more physical picture of the mechanism involved, see ref.~\cite{ESRa}.

Pure $c{=}1$ behaviour is only expected in the very far
infrared. For any finite value of $R$, there will be corrections to
scaling stemming from irrelevant operators compatible with
the symmetry of the model~\cite{Cb}.
Apart from $T\bar T$ and its descendants, the most relevant of
these when the theory is self-dual have conformal dimensions $(N/4,N/4)$.
The model is therefore described by the continuum action
\eq
S=S^*_{\rm IR} +\mu_1\!\int\!\psi_{N/4}\,d^2x
+\mu_2\!\int\! T\bar T\, d^2x + {\rm (further~terms)}~,
\label{IRact}
\en
where $S^*_{\rm IR}$ is the action~(\ref{bosact}), and $\psi_{N/4}$ 
denotes an equal (self-dual) combination of the unit vortex and $N$-fold
symmetry-breaking fields. The further terms are not
optional in the continuum theory: the perturbation theory is 
non-renormalisable, and so infinitely many counterterms are needed.
Nevertheless, it is possible to extract some non-trivial predictions:
an asymptotic expansion for $RE(R)$,
in various powers of $R$ and $\mu_1$,
$\mu_2$, $\dots\,$, can be obtained after a mapping from the cylinder
to the plane~\cite{Zc}.

The fact that $\psi_{N/4}$ is primary means that its contributions
only start at order $\mu_1^2\,$, and the charge conservation selection
rule for vertex operator correlation functions restricts subsequent
terms to even powers. For this operator, $y=(2{-}2\Delta)= 2{-}N/2$,
and so its leading correction to $RE(R)$ is generally proportional to 
$\mu_1^2R^{4{-}N}$.

By contrast, $T\bar T$ is a secondary operator and has a non-vanishing
one-point function on the cylinder. Consequently, despite its
renormalisation group eigenvalue $y$ being equal to $4$, the corrections
due to $T\bar T$ start early, with a term proportional to $\mu_2R^{-2}$.

Evidence for $\psi_{N/4}$ can be found in the
influence of the magnonic pseudoenergies on the large $R$ solutions of the
massless TBA equations, using a small extension of an argument given by
Zamolodchikov~\cite{Zd}. In the $R{\rightarrow}\infty$ limit, the
pseudoenergies acquire the form of a pair of sine-Gordon kink systems, 
one located near $\theta{=}{-\log\MR}$, and one near 
$\theta{=}{\log\MR}$. In the central region between the kinks, 
the non-magnonic pseudoenergies are dominated by their energy terms, and play
no direct r\^ole in the equations:
the functions $L^{\sscr(1)}_n$ 
and $L^{\sscr(\wtilde 1)}_n$, which would otherwise couple to the
reduced system of magnonic pseudoenergies, are doubly-exponentially
suppressed.
Their dominant effect comes instead via
the `tails' of the relevant kernel functions, which only suffer a
single-exponential decay.

Thus for $N=4n{+}2$ (when the TBA system~(\ref{TBAzbn}) applies) the
magnonic pseudoenergy 
$\ep^{\sscr(2)}_n(\theta)$ sees $\ep^{\sscr(1)}_n(\theta)$ and 
$\ep^{\sscr(3)}_n(\theta)$ via the 
convolutions $\phi_1{*}L^{\sscr(1)}_n(\theta)$ and 
$\phi_1{*}L^{\sscr(3)}_n(\theta)$.
In the central region these expand as odd series in $e^{\pm h\theta}$
(recall that $\phi_1(\theta){=}h/\!\cosh h\theta\,$), implying that
$Y^{\sscr(2)}_n(\theta){=}e^{\ep^{\sscr(2)}_n(\theta)}$ 
satisfies the equation
\eq
Y^{(2)}_n(\theta-\fract{\imath\pi}{2h})
Y^{(2)}_n(\theta+\fract{\imath\pi}{2h})=1~.
\label{IRYone}
\en
(Note that this is just the truncation of the full Y-system
to the central magnonic node.)
Therefore $Y^{\sscr(2)}_n(\theta{+}
\fract{2\pi\imath}{h})=Y^{\sscr(2)}_n(\theta)$.
Repeating the argument used in the ultraviolet limit, this
periodicity implies that the leading `magnonic' corrections to $RE(R)$ are 
powers of $(\MR)^{-2h}=(\MR)^{4{-}N}$.

For $N=4n{+}1$ and $N=4n{+}3$, the magnonic pseudoenergies form a
$d_4$-type system with kernel $\phi_2(\theta)=2h/\!\cosh 2h\theta$. The
pseudoenergies $\ep^{\sscr(1)}_n(\theta)$ and $\ep^{\sscr(6)}_n(\theta)$
are felt via $\phi_2{*}L^{\sscr(1)}_n(\theta)$
and $\phi_2{*}L^{\sscr(6)}_n(\theta)$.
These expand in the central region
as odd powers of $e^{\pm 2h\theta}$, and such terms cancel out when
shifted by $\pm\imath\pi/4h$ and summed. As a result they do not impede 
the construction of the Y-systems. For $N=4n{+}1$, the system is
\eq
Y^{(\alpha)}_n(\theta-\fract{\imath\pi}{4h})\,
Y^{(\alpha)}_n(\theta+\fract{\imath\pi}{4h})=
\prod^5_{\beta=2}
\lf(1+Y^{(\beta)}_n(\theta)\ri)^{l^{[d_4]}_{\alpha\beta}}~,
\en
while for $N=4n{+}3\,$,
\eq
Y^{(\alpha)}_n(\theta-\fract{\imath\pi}{4h})\,
Y^{(\alpha)}_n(\theta+\fract{\imath\pi}{4h})=
\prod^5_{\beta=2}
\lf(1+Y^{(\beta)}_n(\theta)^{-1}\ri)^{-l^{[d_4]}_{\alpha\beta}}~.
\label{IRYtwo}
\en
The index $\alpha$ runs from $2$ to $5$ in both cases.
Again, these are just truncations of the full Y-systems.
They imply the periodicity 
$Y^{\sscr(\alpha)}_n(\theta{+}\fract{2\pi\imath}{h})=
Y^{\sscr(\alpha)}_n(\theta)$~\cite{Zf}, after which the argument runs 
just as for $N=4n{+}2$, and again predicts corrections to scaling 
as powers of $(\MR)^{4-N}$.

Finally, to the case $N{=}4n{+}4$. 
There are no magnonic pseudoenergies in the
TBA system, and so all corrections to scaling must come from the `direct'
interaction between the non-magnonic pseudoenergies, to be discussed
shortly. This would seem to be a problem, since the general idea will be
to attribute these corrections to $T\bar T$ and descendants. Two points
can be made. First, contributions from $\psi_{N/4}$, expected to appear as
powers of $(\MR)^{-4n}$, will anyway be badly tangled up with those from
$T\bar T$ and its descendants. However this point applies equally when
$N=4n{+}2$, where at least for $N=6$ the entanglement manifests itself in a
logarithm at order $(\MR)^{-2}$ (see below). Such a logarithm is 
{\it not} found at
order $(\MR)^{-4}$ in the $\ZZ_8$ theory: in fact the
coefficient of this term fits perfectly the hypothesis of a pure $T\bar T$
perturbation. (Note, coefficients up to high order can be obtained for
this case by doubling the (exact) results for the 
tricritical Ising to Ising flow obtained by Zamolodchikov~\cite{Zd}.)
The resolution comes on re-examining the formula~(\ref{bform}) for the 
coefficients in conformal perturbation theory: when $N{=}4n$, 
$\Delta{=}n$ and is a positive
integer, and $B_2$ vanishes. Thus signs of $\psi_{N/4}$ anyway
appear anomalously
late in the series. Without any control of the counterterms, it is not
possible to be any more precise than this, but at least the 
proposal has not turned out to be inconsistent.

So much for $\psi_{N/4}$. For $T\bar T$, the analysis can be pushed a
little further. To eliminate the corrections already discussed, it is
convenient to replace the TBA
equations with a set in which the magnonic pseudoenergies
associated with the two kink systems are separated, by doubling
them up and coupling one copy only to $\ep^{\sscr(1)}_n$, and the other
only to $\ep^{\sscr(\tilde 1)}_n$. Since the magnonic interaction only
cuts in at order $(\MR)^{4-N}$, the approximation should be good at least
up to this point. For $N{=}4n{+}4$, no approximation is needed at all.
The kinks near $\theta{=}{-\log\MR}$ and $\theta{=}{\log\MR}$ now only
interact via the tails of the kernel functions $\psi_{ij}(\theta)$. 
In Zamolodchikov's original paper on the massless TBA~\cite{Zd} an
iterative procedure was used to account for such an interaction.
It turns out that 
in all of the subsequently-introduced TBA systems for the flows
$g^{(2)}{\times}g^{(1)}/g^{(3)}\rightarrow 
g^{(1)}{\times}g^{(1)}/g^{(2)}$, $g\in ade$~\cite{Ma,Ra},
the first two corrections to infrared scaling also have a simple
form.  Only the case $g=d_{n{+}1}$ will 
be directly relevant here, but the general
result will be recorded as it does not seem to have been given
elsewhere. The TBA systems were given in their full generality in
ref.~\cite{Ra}, and involve kernel
functions $\phi_{ij}(\theta)$ and $\psi_{ij}(\theta)$, related to the
corresponding purely elastic S-matrix elements in the way described
for the $d_{n+1}$ case after equation~(\ref{TBAziv}).
Using some tricks previously employed in the
(ultraviolet) calculation of the bulk terms for certain massive flows
from the 
$g^{(1)}{\times}g^{(1)}/g^{(2)}$ cosets~\cite{KMb},
the final result is
\eq
\fract{1}{2\pi}RE(R)=-\fract{1}{12}c(\infty)+{C_1\over
(\MR)^2}+{C_2\over (\MR)^4} + O\!\lf((\MR)^{-6}\ri)
\label{IRexp}
\en
where $c(\infty)$ is the central charge of the infrared limit. The
coefficients $C_1$ and $C_2$ are
\eq
C_1= -{c(\infty)^2\over 12}\,{\pi\over 3}{M^2\over M_1^2}\psi^{(1)}_{11}
\quad,\quad
C_2= -{c(\infty)^3\over 6}\lf({\pi\over 3}{ M^2\over
   M_1^2}\psi^{(1)}_{11}\ri)^2
\label{IRcoeffs}
\en
where $M_1$ is the mass of the lightest particle in the theory, and
$\psi^{(1)}_{11}$ is the first coefficient of an
expansion of the kernel $\psi_{ij}(\theta)\,$:
\[
\psi_{ij}(\theta)=-\sum_{s=1}^{\infty}\psi^{(s)}_{ij}e^{-s|\theta|}~.
\]
These coefficients are only non-zero when $s$ taken modulo $h$ is an 
exponent of the relevant non-affine
algebra $g$. An explicit expression is
\eq
\psi^{(s)}_{ij}=\frac{h}{\sin\fracs{\pi s}{h}}\, q_i^{(s)}q_j^{(s)}~,
\label{qresult}
\en
where $q^{(s)}_i$ and $q^{(s)}_j$ are components of a unit-normalised
eigenvector of the Cartan matrix of $g$ with eigenvalue 
$2{-}2\cos\frac{\pi}{h}s$. This can be derived 
from the general S-matrix formulae of ref.~\cite{Da}, 
using the identity
\eq
\sum_{p=0}^{h{-}1}(\lambda_i,w^{-p}\phi_j)\sin(2p{+}1{+}u_{ij})
\fract{\pi s}{h} = \frac{h\,q^{(s)}_iq^{(s)}_j}{2\sin\frac{\pi s}{h}}
\en
which is the Fourier transform of equation~(2.14) of ref.~\cite{Db}. 
(See refs.~\cite{Da,Db} for an explanation of
the notation used in this last formula.)

The results~(\ref{IRexp}),(\ref{IRcoeffs}) 
can be compared with the perturbative expansion 
of the action $S^*_{\rm IR}+\mu_2\!\int T\bar Td^2x$~\cite{Zc}:
\eq
\fract{1}{2\pi}RE(R)=-\fract{1}{12}c(\infty)
+{(2\pi)^3 \mu_2\over R^2}\lf({c(\infty)\over 24}\ri)^2
-{(2\pi)^6 \mu_2^2\over R^4}\lf({c(\infty)\over 24}\ri)^3
+ O\!\lf({\mu_2^3\over R^6}\ri)~.
\label{IRpexp}
\en
This much of the expansion is uncontaminated by counterterms, and
therefore gives the unambiguous prediction $C_1^2/C_2=-c(\infty)/24$ if 
the TBA is to be compatible with conformal
perturbation theory. It is easy to verify that this does indeed hold
when the coefficients are given by~(\ref{IRcoeffs}). (This was checked 
numerically by Martins, for $g{=}a_2\,,$ in ref.~\cite{Ma}.)
A comparison of the two series also allows the renormalised coupling
$\mu_2$ to be expressed in terms of the crossover scale $M$:
\eq
\mu_2=-{2M^2\over\pi^2M_1^2}\psi^{(1)}_{11}\,M^{-2}.
\en
Note that for any $g\in ade$, this $\mu_2$ is negative: 
for a unitary flow 
with $T\bar T$ contributing the leading infrared correction,
this has to be the case, since otherwise the
expansion~(\ref{IRpexp}) would contradict
Zamolodchikov's $c$-theorem.

Specialising now to $g{=}d_{n{+}1}$, we obtain results relevant to 
the self-dual $\ZZ_{N}$ flows with $N{=}4n{+}4$. For other values of
$N$, the kernel $\psi_{nn}(\theta)$ is more complicated.
Nevertheless, the identity
\[
\psi^{(1)}_{ij}=\frac{M_iM_j}{M_1^2}\psi^{(1)}_{11}\quad (i,j=1\dots n)
\]
(previously a simple consequence of equation~(\ref{qresult})) 
continues to hold and
allows the calculation to go through as before, so that the 
results~(\ref{IRexp}),(\ref{IRcoeffs}) are again recovered. Substituting 
$c(\infty)=1$, $M_1=2M\sin\fract{2\pi}{N{-}4}$, and
$\psi^{(1)}_{11}=4\sin\fract{2\pi}{N{-}4}$, the final values for the
expansion coefficients and $T\bar T$ couplings read
\bea
&&C_1=-\frac{1}{12}\lf(\frac{\pi}{3\sin\frac{2\pi}{N{-}4}}\ri)\quad,\quad
C_2=-\frac{1}{6}\lf(\frac{\pi}{3\sin\frac{2\pi}{N{-}4}}\ri)^2~;\nn\\[3pt]
&&\mu_2=-\frac{2}{\pi^2\sin\frac{2\pi}{N{-}4}}\,M^{-2}\,~.
\label{IRZresults}\\ 
\nn
\eea
\vskip -6pt
These formulae break down when $N{=}5$ and $N{=}6$. This is
not too surprising: the whole derivation fails for $N{\le}
7$, since the direct interaction kernels $\psi_{ij}(\theta)$ are zero in
these cases. 
As the numerical results below show, $1/R^2$ corrections do nonetheless 
appear. At a qualitative level, these can be seen to come,
exceptionally, from the magnonic kernels (though for $N{=}7$ it
is necessary to invoke the thus-far disregarded kernels $\phi_3$,
$\phi_4$ and $\phi_5\,$). However, for exact results it 
is more fruitful to suppose that an analytic continuation of the
formulae valid for larger values of $N$ makes sense. For $N{=}7$ this
is easily done, and gives a result that checks well with the numerical
data. For $N{=}5$ and $N{=}6$, a little more work is needed: 
in the same spirit as for the bulk term logarithms, the `$T\bar T$'
divergences given by~(\ref{IRZresults}) should balance against
divergences in other parts of the expansion, allowing the final result
to remain finite. Indeed, at these two values of $N$ an additional
contribution proportional to $R^{-2}$ comes from the term of 
order $\mu_1^{2/(N{-}4)}$ 
in the perturbative expansion, and this has the
potential to cancel the pole in $C_1$ (furthermore,
the regularised $B_2$ itself diverges when $N{=}6$).
To find the residual finite term, set $m=2/(N{-}4)$ and evaluate
\eq
\lim_{N\rightarrow 4+{2\over m}}\lf[
\frac{-1}{12}\,\frac{\pi}{3\sin\frac{2\pi}{N{-}4}}
\lf(\frac{1}{(\MR)^2}-\frac{1}{(\MR)^{(N{-}4)m}}\ri)\ri]
= {(N{-}4)\over 36\cos\fract{2\pi}{N{-}4}}\,{\log\MR\over(\MR)^2}~.
\label{logcalc}
\en
This gives the corrected coefficients
\eq
C_1^{(N{=}5)}=\frac{1}{36}\log\MR+{\rm const}\quad,\quad
C_1^{(N{=}6)}=-\frac{1}{18}\log\MR+{\rm const}~.
\label{Cilogs}
\en
These values can also be extracted directly from the
TBA, as explained in section~4.2 below.
To deal with 
$C_2$, {\it two} additional terms must be taken into account, at
perturbative orders $\mu_1^{2m}$ and $\mu_1^m\mu_2\,$: both become
proportional to $R^{-4}$ when $N=5$ or $6$. In terms of
$x=N-(4{+}\fract{2}{m})$, they should cancel the $x{\rightarrow}0$
divergences in 
\[
C_2(x)=-\frac{2}{27m^4}
\lf[\frac{1}{x^2}+\frac{m}{x}+\frac{3m^2}{2}+\frac{m^4\pi^2}{12}+\dots\ri]~.
\]
To achieve this, $a(x)$ and $b(x)$ must be found such that
\[
\frac{1}{x^2}\lf(\frac{1}{(\MR)^4}+
\frac{mx}{(\MR)^4}
- \frac{a(x)}{(\MR)^{4+2mx}}-
\frac{b(x)}{(\MR)^{4+mx}}\ri)
\]
is finite at $x{=}0$, determining $a(x)=-1+\alpha x+O(x^2)\,$,
$~b(x)=2+(m{-}\alpha)x+O(x^2)$. In turn this fixes that part of the
finite residue proportional to $\log^2\MR$. The upshot is that the
earlier expression for $C_2$ should be  modified to
\eq
C_2^{(N{=}5{,}6)}=-\frac{(N{-}4)^2}{54}\log^2\MR +{\rm const}.\log \MR
+{\rm const}'~.
\label{Ciilogs}
\en
Infrared logarithms appearing to a single power
have previously been observed numerically, in
the interpolating flows
between the low-lying minimal models ${\cal{M}}_p\rightarrow
{\cal{M}}_{p-1}$~\cite{KMc,FQRa}, and also analytically, in a massless sausage
model~\cite{FOZa}. The method used in section~4.2
to extract the $\ZZ_N$ logarithms directly from the TBA can 
also be applied to these other flows, and the calculations are
described in second half of that section.

\bigskip\noindent
We also investigated the infrared behaviour of the TBA systems numerically,
both to verify the results derived above and to provide information on
coefficients that we have not been able to extract exactly. 
For $R$ ranging from $120$ to $11000$,
$\fracs{1}{2\pi}RE(R)$ was found for $N=5$, $6$ and $7$.
The iterative procedure was as described in the last section, and again the 
normalisation $M{=}1$ was adopted. 

First, we estimated the exponents of the leading correction
terms by finding the limiting slopes of  plots of
$\log(\fracs{1}{2\pi}RE(R)+\frac{1}{12})$ against $\log R$, 
with the results
$-0.9993$, $-1.9967$ and $-2.0010$ respectively (to be compared with
the predictions of $-1$, $-2$ and $-2$). 
Thus reassured that at least the leading behaviour was
as expected, we then fitted the data
to expansions in $R$ and $\log R$ of the predicted forms,
leaving all the coefficients unconstrained. 
Numerical precision was not as good as in the ultraviolet, 
possibly reflecting the asymptotic nature of the infrared 
expansions. The following fits were obtained 
for $\fracs{1}{2\pi}RE(R)+\frac{1}{12}~$:
\bea
N{=}5~&:&~
- \frac{0.0177380}{R} + \frac{0.027733 \log R }{R^{2}}
 -\frac{0.01951}{R^{2}} +\frac{0.097}{R^3} +\dots \nn\\[3pt]
N{=}6~&:&~
 - \frac{0.05555546 \log R }{R^{2}}
+\frac{0.033585}{R^{2}}
-\frac{0.0715\log^2R}{R^4}+\dots\nn\\[3pt]
N{=}7~&:&~
 - \frac{0.1007662}{R^{2}} + \frac{0.153}{R^{3}} +\dots\\\nn
\eea
\vskip -4pt \noindent
In each case, the omitted constant piece (measuring the difference between
the numerical and 
exact  values for $c(\infty)/12\,$) was zero to 11 significant figures.
For $N{=}5$ and $6$, the coefficients of $R^{-2}\log R$ match well with
the predicted values of $0.02777\!\dots\,$ and $-0.05555\!\dots\,$
respectively, while for $N{=}7$ the formula~(\ref{IRZresults})
predicts the coefficient of $R^{-2}$ to be $-0.1007663\!\dots\,$, 
again in good agreement with the numerical results.
The other predictions are more difficult to verify: if the 
expansions really are
asymptotic, then results at larger and larger values of $R$ are needed 
before higher terms can be captured, and one rapidly runs up against
the limitations in numerical accuracy.
For $N{=}7$, a coefficient of $0.24369\!\dots\,$ is predicted for the
$R^{-4}$ term, while for $N{=}5$ and $N{=}6$
the coefficients of $R^{-4}\log^2R$  should 
be $-0.0185185\!\dots$ and $-0.074074\!\dots$ respectively. 
The agreement for $N{=}6$ is reasonable, and improved when a
restricted fit (using the exact values for $c(\infty)$ and $C_1$) was
performed, the estimate for this coefficient changing to
$-0.0743\dots\,$.
For the other cases, no such convergence was observed for any
coefficients beyond those reported above.
This is not particularly surprising, since for $N{=}5$
there are more unknowns to fix before the
term of interest is reached, and for $N{=}7$ it is anyway smaller,
lacking the $\log^2R$ factor. 
Greater numerical precision will be needed before any more can be
said.

Nevertheless, in all instances where we have reliable numerical 
results, they agree with the exact predictions. Since some of these 
latter were derived using an additional physical assumption of
analyticity in $N$, this lends
further credibility to the whole scenario.

One further remark can be made before leaving the Kosterlitz-Thouless
phase. If
the action defined by (\ref{IRact}) and (\ref{bosact}) is converted
into the normalisations of appendix~A by setting
$\Phi=\sqrt{\pi}\varphi$, then the $N$-fold symmetry breaking part of
the perturbation becomes proportional to $\cos(\sqrt{2N\pi}\,\varphi)$,
and formally matches the sine-Gordon Lagrangian~(\ref{SGact}) at
$\beta^2=2N\pi$. Toda-type duality (see for example ref.~\cite{BCDSa}) 
maps $\beta^2$ to $64\pi^2/\beta^2$ in these normalisations. Thus
$\beta^2=2N\pi$ is Toda-dual to $\beta^2=32\pi/N$, the value already
picked out in section~2. This gives an
alternative perspective on these special values of the sine-Gordon
coupling constant.

\subsection{Exact infrared information from the TBA}
This section, something of an interlude from the main development
of the paper, discusses some general properties of massless TBA
systems. Motivated by the results obtained above, the particular
interest is the mechanism by which 
logarithmic terms can be generated at large $R$. 
The simplest example of this phenomenon is
the flow between the minimal models ${\cal{M}}_5$ and 
${\cal{M}}_{4}$~\cite{KMc}. The relevant
TBA system, proposed by Zamolodchikov~\cite{Zd}, reads
\bea
\halft\MR e^{\theta}&=&\ep_1(\theta)+\phi{*}L_2(\theta)
\label{one}\\
0&=&\ep_2(\theta)+\phi{*}L_1(\theta)+\phi{*}L_3(\theta)
\label{two}\\
\halft\MR e^{-\theta}&=&\ep_3(\theta)+\phi{*}L_2(\theta)
\label{three}\\ \nn
\eea
\vskip -12pt\noindent
with $\phi(\theta)=1/\!\cosh\theta$, $L_i=\log(1+e^{-\ep_i})$, and
\bea
\frac{1}{2\pi}RE(R)=-\frac{1}{12}c(R)
&=&-\frac{1}{4\pi^2}\iint\!d\theta\MR e^{\theta}L_1(\theta)\nn\\
&=&-\frac{1}{4\pi^2}\iint\!d\theta\MR e^{-\theta}L_3(\theta)\,.\nn\\ \nn
\eea
\vskip -5pt\noindent
Take the derivative with respect to $\theta$ of~(\ref{one}), multiply by
$-\frac{1}{2\pi^2}L_1(\theta)$, and integrate from $-\infty$ to $\infty$ 
to find
\eq
\frac{1}{2\pi}RE(R)=
-\frac{1}{2\pi^2}\iint\!d\theta\,\dth\ep_1(\theta)L_1(\theta)
-\frac{1}{2\pi^2}\iint\!d\theta\lf(\phi{*}\dth L_2(\theta)\ri)L_1(\theta)\,.
\label{four}
\en
At $R{=}\infty$, $\ep_1$ solves a kink system obtained from 
equations~(\ref{one})--(\ref{three}) by omitting all 
terms involving $\ep_3$, and vice versa for $\ep_3$. In line with the
notation used earlier in the paper, 
write $\epR_1$, $\epR_2$ for the solutions to the first
kink system, and 
$\epL_2$, $\epL_3$ for the solutions to the second.
To find the logarithmic correction, which comes from an integral over
the central region $-\log\MR\ll\theta'\ll\log\MR$ between the two sets
of kinks, replace $L_1$ with $\LR_1$ and $\dth L_2$ 
by $\dth\LR_2+\dth\LL_2$. This gives
\eq
\frac{1}{2\pi}RE(R)=-\frac{1}{12}c(\infty)
-\frac{1}{2\pi^2}\iint
\!d\theta\lf(\phi{*}\dth\LL_2(\theta)\ri)\LR_1(\theta)~.
\label{five}
\en
In the central region $\epL_2(\theta')$
satisfies
\bea
0&=&\epL_2(\theta')+\frac{1}{2\pi}\iint\!d\theta''
\frac{1}{\cosh(\theta'-\theta'')}\LL_3(\theta'') \nn \\
&=&\epL_2(\theta')+\frac{e^{\theta'}}{\pi}\iint\!d\theta''\,
e^{-\theta''}\LL_3(\theta'')+\dots\nn \\
&=&\epL_2(\theta')+\frac{\pi c(\infty)}{3\MR}e^{\theta'}+\dots\,,
\label{six} \\ \nn
\eea
\vskip -12pt\noindent
and so
\eq
\dth\LL_2(\theta')=\frac{\pi c(\infty)}{6\MR}e^{\theta'}+\dots\,.
\label{seven}
\en
Because of the double-exponential suppression
of $\LR_1(\theta)$ in the central region, 
$\phi(\theta{-}\theta')\dth\LL_2(\theta')$ can be replaced there by
$2e^{\theta-\theta'}\dth\LL_2(\theta')$. It therefore
has a piece constant in
$\theta'$, equal to $e^{\theta}\pi c(\infty)/3\MR$.
Dividing by $2\pi$ and integrating across the central 
region for the
convolution then produces a factor $\frac{1}{\pi}\log\MR\,$, and 
equation~(\ref{five})
becomes
\bea
\frac{1}{2\pi}RE(R)&=&
-\frac{1}{12}c(\infty)-\frac{1}{2\pi^2}\iint\!d\theta 
e^{\theta}\frac{c(\infty)}{3\MR}
\log\MR\,\LR_1(\theta)+\dots\nn\\[2pt]
&=&
-\frac{1}{12}c(\infty)-\frac{c(\infty)^2}{18(\MR)^2}\log\MR+\dots\,.
\label{eight} \\
\nn\eea
\vskip -5pt\noindent
With $c(\infty)=7/10$, the coefficient of the logarithmic term is 
$-49/1800=-0.027222\dots$, which compares well with the value $-0.02723(2)$
found numerically by Klassen and Melzer in ref.~\cite{KMc}.

The calculation goes through essentially unchanged if the TBA system is
augmented by equal numbers of rightmoving and leftmoving pseudoenergies,
so long as they interact only with $\ep_1$ and $\ep_3$ respectively. Their
effect is to change the value of $c(\infty)$, and the
result~(\ref{eight}) remains valid. In particular, the $\ZZ_6$ TBA system, a
special case of~(\ref{TBAzbn}), can be obtained in this way, simply by adding
two magnonic pseudoenergies $\ep_0$ and $\ep_4$ to the 
system~(\ref{one}-\ref{three}). (Note, when $N{=}6$, $\phi_1(\theta)$ is
exceptionally equal to $1/\!\cosh\theta\,$.) Since $c(\infty){=}1$ for this
system, the value of $\Cam{N=6}_1$ given by equation~(\ref{Cilogs}) follows
immediately.

A different generalisation is needed to find $\Cam{N=5}_1$, since for
$N{=}5$ the left and right kink systems influence each other via a reduced
system containing more than one magnonic pseudoenergy. This will modify the
result~(\ref{seven}), but if $\dth\LL_2(\theta')$ retains a piece
proportional to $e^{\theta'}$ 
in the central region, a logarithm will still appear. Formally
the question is the same as
arises in Zamolodchikov's treatment of ultraviolet logarithmic terms, 
described in appendix~A of ref.~\cite{Zd}, since to leading
order the equations 
are the same as those for a standard
TBA. The one change is that instead of energy terms 
$\half\MR e^{\pm\theta}$, the reduced system is driven by terms
$\frac{\pi c(\infty)}{3\MR}e^{\pm\theta}$, 
appearing for those pseudoenergies
which in the full system would be coupled to the remaining right or left
moving pseudoenergies. 
These are induced in the same manner as already seen in
equation~(\ref{six}). (An analogous effect was important in the 
derivation of equations~(\ref{IRYone})--(\ref{IRYtwo}) of section~4.)
By reversing some of the steps of Zamolodchikov's ultraviolet
argument, and then repeating them in the new context, 
it is possible to derive the following prescription. 
Suppose that the TBA based on the reduced set of 
pseudoenergies would have shown an ultraviolet logarithmic
correction to $\frac{1}{2\pi}RE(R)$ equal to 
$\Flog(\MR/2\pi)^2\log\MR$, had the energy
terms been of the usual form. 
Then towards the infrared limit of the full problem, 
$\dth\LL_2(\theta')$ develops a piece proportional
to $e^{\theta'}$ in the central region:
\[
\dth\LL_2(\theta')=\dots + \Flog\frac{\pi c(\infty)}{3\MR}e^{\theta'}
+\dots\,\qquad(-\log\MR\ll\theta'\ll\log\MR )\,.
\]
Note, in contrast to 
equation~(\ref{seven}), this will not necessarily be the
leading term. Nevertheless it gives the leading logarithmic contribution, 
feeding through the remainder of the calculation to predict
\[
\frac{1}{2\pi}RE(R)=
-\frac{1}{12}c(\infty)+\dots+
\frac{\wtilde C_1}{(\MR)^2} \log\MR+\dots\,, 
\]
\eq
\llap{with\qquad\qquad\qquad}\wtilde C_1=
\frac{1}{9}\Flog c(\infty)^2~.\qquad\qquad\qquad\qquad  \label{blob} 
\en
The earlier result~(\ref{eight}) can be recovered on noting that 
the reduced TBA there is the 
same as that of the thermally-perturbed Ising model, for which
$\Flog=-1/2$. For the $\ZZ_5$ TBA, the reduced
system is based on the $d_4$ Dynkin diagram, with driving terms
attached one to each node of the fork, and 
$\Flog=1/4$~\cite{FZb}. Hence
$\CTam{N=5}_1 =1/36\,$, as previously
obtained by other means.

The same method can be applied to the sausage models
$SST_{\lambda}^{(-)}$ at $\lambda=1/K$, $K=3,4,\dots\,$, using the TBA
systems given in ref.~\cite{FOZa}. 
These models arrive at $c(\infty){=}1$, attracted in the infrared 
by an operator with conformal dimension
$\Delta_{IR}=1/(1{-}\lambda)=K/(K{-}1)$.
The reduced system is based on the
(non-affine) $d_{K{-}1}$ Dynkin diagram, with (for $K\ge 4$)
both left and right
moving driving terms coupling to the extremal node of the tail. This
entails $\Flog=0$ for $K$ even, and
$\Flog=-2/(K{-}1)$ for $K$ odd~\cite{RTVa}. Hence
\eq
\CTam{\lambda=1/K}_1 
= \left\{ \begin{array}{ll}
\displaystyle 0 & ~~(K \mbox{ even})\nn\\[2pt]
\displaystyle\frac{-2\m}{9(K{-}1)} & ~~(K \mbox{ odd})\nn\\
\end{array} \right.
\label{sauslog}
\en
When $K{=}3$ the reduced system is disconnected, and there are two
contributions to the logarithmic term. When these are added together,
the general formula turns out to be valid in this case as well. In
ref.~\cite{FOZa} the same result for $K{=}3$ was derived by an
indirect route, closer to the analytic continuation argument used
elsewhere in this paper. Part of the input was the pair of formulae
\eq
\Cam{\lambda}_1=-\frac{\pi}{18}\tan\frac{\pi}{2\lambda}\quad,\quad
\mu^{\sscr(\lambda)}_2=-\frac{4}{\pi^2}\tan\frac{\pi}{2\lambda}M^{-2}~,
\label{FOZres}
\en
found by putting the model in an external magnetic field. 
In fact this information is enough to predict the coefficient of the
logarithm for all integer $K$, with agreement with the direct TBA
derivation. This gives a further check on the results of ref.~\cite{FOZa}.
Going further (and contrary to equation~(5.47) of~\cite{FOZa}), arguments 
similar to those used in section~4 predict an additional
$\log^2\MR/(\MR)^4$ correction for $K$ odd, with 
coefficient $-32/27(K{-}1)^2$. This was checked numerically for $K{=}3$.

One other special case of the general result~(\ref{blob}) is of some
interest. This covers the TBA systems proposed by Zamolodchikov in
ref.~\cite{Ze} to describe the flows
\eq
\CM A^{(+)}(k,l)~:~~ \CM(k,l)\rightarrow\CM(k,l{-}k)
\label{leE}
\en
from the coset models $\CM(k,l)\equiv
a_1^{(k)}{\times}a_1^{(l)}/a_1^{(k{+}l)}$. The pseudoenergies 
live on an $a_{l+k-1}$ Dynkin diagram, with a rightmoving energy term
$\halfs\MR e^{\theta}$ on the $k^{\rm th}$ node, and a symmetrically-placed
leftmoving term
$\halfs\MR e^{-\theta}$ on the $l^{\rm th}$ node.
The reduced system operating in
the central region sees the $a_{l-k-1}$ Dynkin diagram, with the
induced energy terms sitting one at each end. If $l{-}k{-}1$ is odd then the
ultraviolet version of such a system shows a logarithmic singularity, with
$\Flog=2(-1)^{(l{-}k)/2}/(l{-}k{+}2)$~\cite{Zd}. 
Therefore the flow~(\ref{leE}) will show a logarithm in the infrared
whenever $l{-}k$ is even. The coefficient is given by~(\ref{blob}),
with
$c(\infty)=c(k,l{-}k)=3k(l{-}k)(l{+}4)/(k{+}2)(l{+}2)(l{-}k{+}2)\,$.
Setting $k{=}2$, $l{=}4$ recovers the $\ZZ_6$ result, while
the cases $k{=}1$, $l{=}p{-}2$
pertain to the flows $\CM A^{(+)}_p$: $\CM_p\rightarrow\CM_{p-1}$
between minimal models. Therefore, for
these flows
\eq
\CTam{p}_1 
= \left\{ \begin{array}{ll}
\displaystyle 0 & ~~(p \mbox{ even})\nn\\[2pt]
\displaystyle (-1)^{(p{-}3)/2}\,\frac{2c(\infty)^2}{9(p{-}1)}
& ~~(p \mbox{ odd})\nn\\
\end{array} \right.
\label{minlog}
\en
where $c(\infty)=1-6/p(p{-}1)\,$. An analogous calculation
for the massless flows $H^{(\pi)}_{p-2}$:
$Z_{p-2}\rightarrow\CM_{p-1}$ introduced in ref.~\cite{FZb} 
(where $Z_{p-2}$ denotes
the $\ZZ_{p-2}$-symmetric conformal field theory) finds the same
formula (\ref{minlog}), save for the replacement of the prefactor
$(-1)^{(p-3)/2}$ by $-1$. The similarity of the $H^{(\pi)}_{p-2}$
result to the
sausage formula (\ref{sauslog}) is not coincidental. 
As recently stressed in ref.~\cite{FQRa}, it appears to be consistent
to regard the theory $H^{(\pi)}_{p-2}$ as a reduction of the sausage
model $SST^{(-)}_{1/p}$. Furthermore, in such a case the ground-state
energy $E^{\rm r}(R)$ after reduction should be equal to
that of a suitable excited state $E^{\rm u}_k(R)$ in the unreduced 
model (this
is clearly explained for the reductions of the sine-Gordon model in
ref.~\cite{Zi}). For the reduced ultraviolet and infrared central 
charges to emerge correctly from this man\oe uvre, the excited state
must have ultraviolet and infrared scaling dimensions 
\[
2\Delta_k(0)=
\fract{1}{12}\lf(c^{\rm u}(0)-c^{\rm r}(0)\ri)\quad,\quad
2\Delta_k(\infty)=
\fract{1}{12}\lf(c^{\rm u}(\infty)-c^{\rm r}(\infty)\ri)~,
\]
with $c^{\rm u}(0)$, $c^{\rm u}(\infty)$ and $c^{\rm r}(0)$,
$c^{\rm r}(\infty)$ the ultraviolet and infrared central charges 
before and after reduction. In infrared perturbation theory, this is 
reflected in the modification to the $T\bar T$ corrections to
scaling for an excited state with infrared scaling dimension 
$2\Delta_k(\infty)$~\cite{KMc}: 
the result~(\ref{IRpexp}) still holds, but with
$c^{\rm u}(\infty)$ replaced by 
$c^{\rm u}(\infty){-}24\Delta_k(\infty)=c^{\rm r}(\infty)\,$. Comparing
the $T\bar T$ expansions for $\frac{1}{2\pi}RE^{\rm u}(R)$ and
$\frac{1}{2\pi}RE^{\rm u}_k(R)\,$, this implies that the coefficient
of $(\MR)^{-2}$ for the excited state can be obtained simply 
by multiplying the coefficient for the ground state, $C_1^{\rm u}$,
by $(c^{\rm r}(\infty)/c^{\rm u}(\infty))^2$.
So long as the mass scale remains the same after reduction,
this means that
\eq
C_1^{\rm r}=\lf(\frac{c^{\rm r}(\infty)}{c^{\rm u}(\infty)}\ri)^2
C_1^{\rm u}~.
\label{redres}
\en
A similar result relates $C_2^{\rm r}$ to $C_2^{\rm u}$. 
Resonances may ocassionally cause these coefficients to diverge.
However, so long as the resulting logarithms can be obtained by analytic 
continuation from non-resonant points, their coefficients must
be related in the same way:
\eq
\wtilde C_1^{\rm r}
=\lf(\frac{c^{\rm r}(\infty)}{c^{\rm u}(\infty)}\ri)^2
\wtilde C_1^{\rm u}~.
\label{lredres}
\en
This explains the connection observed above between the logarithmic
singularities for $SST^{(-)}_{1/p}$ and $H^{(\pi)}_{p-2}$ when $p$ is
odd. Furthermore, combining (\ref{FOZres}) with (\ref{redres}) yields
a prediction for $H^{(\pi)}_{p-2}$ when $p$ is even, namely
$C_1^{(p)}=0$.

The flow $\CM A^{(+)}_p$ can also be found as a reduction of a theory
that flows in the infrared to $c{=}1$, this time an imaginary coupled
sine-Gordon model (ISG)~\cite{FSZa}. The flow to take leaves
$c{=}1$ by a relevant operator with conformal dimension $\Delta_{UV}
=p/(p{+}1)$, wanders around in a non-unitary way, 
and then returns to $c{=}1$ via an irrelevant operator of conformal 
dimension $\Delta_{IR}=p/(p{-}1)$ (as usual,
there will also be a collection of counterterms).
Such models can be defined for all $p{\ge}2$.  Unfortunately, 
we do not know very much about their infrared behaviour,
and so we resort instead to the following trick, which finally brings
the discussion back to the TBA systems introduced in section~2.
First, parametrise the arriving dimension $\Delta_N{=}N/4$ for the
self-dual $\ZZ_N$ flow as $\Delta_N=p/(p{-}1)$, and then rewrite 
the formula~(\ref{IRZresults})
in terms of $p\,$: $C_1^{\sscr(N)} =-\pi/(36\sin\frac{\pi}{2}(p{-}1))\,$.
This is singular when $p$ is an odd integer -- precisely the values
where the corresponding coefficient in the ISG should diverge if its
reduction is to reproduce the behaviour of $\CM A^{(+)}_p\,$.
The coefficient of the logarithm arising at these points is, from
equation~(\ref{logcalc}), equal to 
$(-1)^{(p{-}1)/2}/9(p{-}1)\,$.
Comparing with the result~(\ref{minlog}) for $\CM A^{(+)}_p$, $p$ odd, 
and recalling the relation~(\ref{lredres}), leads to the
conjecture that the coefficient $C_1$ for the massless sine-Gordon model 
is minus twice the analytic continuation of the $\ZZ_N$ value: 
\eq
C_1^{\rm (ISG)}=\frac{\pi}{18\sin\frac{\pi}{2}(p{-}1)}~.
\label{ISGresult}
\en
We have only checked this directly for one case: in ref.~\cite{FSZa},
a TBA system was given for the $p{=}2$ ISG, which arrives at $c{=}1$
along operators with the same dimensions as those for the
self-dual $\ZZ_8$ flow. From this system, it is
possible to extract $C_1=\pi/18$, which is indeed the
value given by~(\ref{ISGresult}). Note that the expected monstrous
corrections to the ISG TBA, investigated in ref.~\cite{Zj}, have an
exponentially small influence in the far infrared and so do not change
this result.
At other values of $p$, the conjecture 
leads to a prediction which can at least be checked numerically: for
$\CM A^{(+)}_p$ with $p$ even, a case previously inaccessible to us,
we expect that
\eq
C_1^{\sscr(p)}
=\frac{\pi c(\infty)^2}{18\sin\frac{\pi}{2}(p{-}1)}\qquad (p~{\rm
even})\,.
\en
For $p{=}4$ this reproduces the value 
$C_1^{\sscr(4)}=-\pi/72$ obtained by Zamolodchikov for the tricritical
Ising to Ising flow~\cite{Zd}. 
To go further, we extracted values of $C_1$ or
$\wtilde C_1$ from fits to numerical solutions of the relevant TBA systems,
allowing for the possibility of $\log^2 \MR/(\MR)^4$ terms for $p$ odd, 
for $p=5\dots 10$. The results are compared with the exact predictions in
tables~\ref{tt} and~\ref{tt1}. (The first numerical entry in table~\ref{tt}
was taken from ref.~\cite{KMc}.) 

The paper~\cite{FQRa}, by Feverati {\it et al}, appeared 
as this paper was being written. Amongst other things, it contains 
a collection of further numerical results, in particular for 
the $H^{(\pi)}_{p-2}$ flows, which are also consistent with the exact
results obtained in this section.

\begin{table}
\begin{center}
\vskip -5pt  
\begin{tabular}{|c|l|l|}     \hline \hline
\rll
{}~$M_{p} \rightarrow~M_{p-1}${} &~~~~exact~~&~~~~numeric~~~\\ 
\hline
$M_5\rightarrow  M_4$ & $~  -0.027222\dots~~$ & $~  -0.02723~~$\\ \hline
$M_7\rightarrow  M_6$ & $~\m 0.027211\dots~~$ & $~\m 0.027211 ~~$\\ \hline
$M_9\rightarrow  M_8$ & $~  -0.023341\dots~~$ & $~  -0.0232   ~~$ 
\\\hline \hline \end{tabular}
\end{center}
\vskip -10pt
\caption{ $\wtilde{C}_1$ for the flows $\CM A^{(+)}_p$, $p$ odd.}
\label{tt}
\end{table}
%
\begin{table}
\begin{center}
\vskip -5pt  
\begin{tabular}{|c|l|l|}     \hline \hline
\rll
{}~$M_{p} \rightarrow~M_{p-1}${} &~~~~exact~~&~~~~numeric   ~~~\\ 
\hline
$M_6\rightarrow  M_5$ & $~\m 0.111701\dots~~$ & $~\m 0.11166~~$\\ \hline
$M_8\rightarrow  M_7$ & $~  -0.139137\dots~~$ & $~  -0.13910~~$\\ \hline 
$M_{10}\rightarrow M_9$ & $~\m 0.152038\dots~~$ & $~\m 0.151~~$ 
\\\hline \hline \end{tabular}
\end{center}
\vskip -10pt
\caption{ $C_1$ for the flows $\CM A^{(+)}_p$, $p$ even.}
\label{tt1}
\end{table}

\subsection{Phase coexistence in the massive regime}
For the massive flow, a very different picture is expected 
at large distances.
The ultraviolet perturbing operator is self-dual, 
and so in the massive direction it should move the 
model onto a surface of first-order transitions.
In this region the ordered and disordered phases coexist, and so the
possible vacua should 
form a $\ZZ_N$-multiplet of $N$ ordered ground states,
$\VEV{\sigma}=e^{\fracs{2\pi\imath}{N}k}$, and a disordered
$\ZZ_N$-singlet with $\VEV{\sigma}=0$.

Finite-size scaling at a first-order transition is reasonably
well-understood. In a cylindrical geometry, dominant configurations
consist of a sequence of domain walls stretching across the
`spacelike' dimensions of the system. These lift the ground-state 
degeneracy, replacing it with an energy splitting which in $d{+}1$ 
dimensions is of order $R^{d/2-1}\exp(-\sigma R^d)$~\cite{PFa}. Here
$R^d$ is the spatial volume and $\sigma$ the surface tension of the
domain wall; in $1{+}1$ dimensions $\sigma$ is equal to the mass of
the corresponding single kink.

When there are $m>2$ different infinite-volume ground states, the
standard
discussion must be generalised a little. Consider first the situation
where all kinks have the same mass, the vacua that they connect being
encoded in an incidence matrix $I_{ab}$. Then the usual
instanton-gas argument (see for example ref.~\cite{Cc}) must be
supplemented by a diagonalisation of $I$. The prefactors of $\mp 1$ --
minus the eigenvalues of $\lf({0~1\atop 1~0}\ri)$
 -- which distinguish the energies of the lowest-energy symmetric and
antisymmetric states in the Ising model are replaced by
$-\lambda_0<-\lambda_1\le\dots\le -\lambda_{m-1}$, the negatives of
the $m$ eigenvalues of $I$. (This can equivalently
 be understood via the effective
quantum-mechanical problem that remains once all transverse degrees of
freedom have been integrated out.)
In particular, the negative of the
(unique) largest eigenvalue controls the leading behaviour of the
ground-state energy~\cite{Zc}. Note however that {\it all} of the
eigenvalues will be seen in the full spectrum, the others
appearing in the asymptotics of those 
energy levels which become degenerate with the ground state in
infinite volume. This explains the observation of one instance of this
phenomenon made in ref.~\cite{RSTa}, and it offers the
interesting prospect that it might be possible to `hear' the shape of
the kink structure in the large-$R$ asymptotics of the degenerating
energy levels. We checked that the excited-state TBA systems
reported in refs.~\cite{Mc,KMc} are consistent with this
idea.

The above assumed that all the kinks had the same mass. In general 
there will be $n{\ge}1$ different kinds of kinks, with masses that 
can be ordered as
$M_{\pi(1)}{<}M_{\pi(2)}{<}\dots {<}M_{\pi(n)}$ for some permutation $\pi$ of
$1\dots n$. 
(In the $\ZN$ models, $\pi(1)=n$ for $N{\le}16$, and $\pi(1)=1$ thereafter.) 
For each mass $M_i$, an incidence
matrix $I^{\sscr(i)}_{ab}$ can be defined to encode 
the number of kinks of that mass
which join each pair of vacua. 
If these matrices commute, then they can be simultaneously
diagonalised to give sets of simultaneous eigenvalues
\eq
(\lambda^{(1)}_0,\lambda^{(2)}_0,\dots\lambda^{(n)}_0) >
(\lambda^{(1)}_1,\lambda^{(2)}_1,\dots\lambda^{(n)}_1)\ge\dots\ge
(\lambda^{(1)}_{m{-}1},\lambda^{(2)}_{m{-}1},\dots\lambda^{(n)}_{m{-}1})\,.
\label{lambdaorder}
\en
The ordering implied here is lexicographic, first by the eigenvalues 
$\lambda^{(\pi(1))}_p$ 
for the lowest kink mass, then by those for the second-lowest mass, and so 
on. 
The asymptotic behaviour of the $k^{\rm th}$ energy level is then
\eq
E_k(R)\sim -\lambda^{(1)}_k\Lambda_1(R)-
\lambda^{(2)}_k\Lambda_2(R)-\dots
-\lambda^{(n)}_k\Lambda_n(R)~,\quad k=0,1,\dots m{-}1
\label{Easympt}
\en
where $\Lambda_i(R)$, a function of order $R^{-1/2}\exp(-M_iR)$, gives
the leading energy splitting for a system with just two vacua and
kink mass $M_i$.
This formula must be interpreted with some caution: if some of the
heavier kinks have a mass more than twice that of the lightest, then
their leading contributions, included above, will be less important
than sub-leading contributions from the lightest kink, which have been
omitted. Nevertheless, the functional forms of the $\Lambda_i(R)$ provide 
clear fingerprints for leading one-kink terms of each type, and match
with the terms which emerge at the first step of
an iterative solution of the TBA equations. Thus while not
always the leading asymptotic, (\ref{Easympt}) isolates exactly those
terms that will be important below. 

The result seems to be consistent with a previously-known case.
In ref.~\cite{Fc}, excited-state TBA systems
were proposed for the perturbation of the $\ZN$-symmetric
conformal field theory by its first thermal operator $\epsilon^{(1)}$.
In contrast to the $\epsilon^{(2)}$ perturbation that has been the
principal concern of this paper, $\epsilon^{(1)}$ is anti-self-dual and 
moves the
model off the self-dual hyperplane. In the low-temperature direction,
this leads to a theory of kink scattering.
For current purposes it is convenient to group the kinks and antikinks
together, even though their scattering is in fact diagonal, so that
the model exhibits $[N/2]$ different kink types, with masses
$M_j=M\sin(\pi j/N)$,
$j=1\dots [N/2]$. A number of different excited-state energies
were found in ref.~\cite{Fc}; for those with conformal dimensions
$k(N{-}k)/2N(N{+}2)$ at short distances, we extracted the large $R$
asymptotics in the way shortly to be described for the self-dual
flows. The resulting eigenvalues were $\lambda^{(j)}_k{=}2\cos(2\pi
jk/N)$, excepting the self-conjugate $(N/2)^{\rm th}$ kink for $N$
even, for which $\lambda^{(N/2)}_k{=}\cos(\pi k)$. If it is assumed that
these describe the excited states for untwisted boundary conditions 
in the low-temperature phase, then a compatible
set of incidence matrices $I^{(j)}_{ab}$ can be found on setting
$I^{(1)}_{ab}=[\widehat A_{N-1}]_{ab}$, the incidence matrix of the affine
$a^{(1)}_{N-1}$ Dynkin diagram, and then defining $I^{(2)}_{ab}\dots
I^{([N/2])}_{ab}$ via an $SU(2)$-type fusion hierarchy:
\eq
I^{(j)}I^{(1)}=I^{(j{-}1)}+I^{(j{+}1)}~,
\label{fuse}
\en
with $I^{(0)}$ set equal to $2\II$, twice the identity matrix. The
factor $2$ can be traced to the grouping of kink with antikink, which as
before leads to an exceptional case for $N$ even: the matrix
$I^{(N/2)}$ obtained from (\ref{fuse}) must be halved. All of this is
in agreement with the usual picture of the kink structure in the
ordered phase, and in particular matches the bosonic parts of the
supersymmetric solitons studied in~\cite{FMVWa}.

There is a potential problem in more general cases:
the assumption that the incidence matrices commute is rather strong,
and one can envisage theories of kink scattering for which it fails. 
(For the example just described, it followed from
the $\ZN$ symmetry of the $N$ degenerate vacua.)
In the related context of integrable lattice models, the property
is often built in from the start
(see for example ref.~\cite{Zg}). However, from the
S-matrix perspective it turns out to be a simple consequence of
integrability, as shown by the following argument.

Consider the model on the full line, with spatial coordinate $x$, and
restrict attention to the sector of those states
which interpolate between the vacua $a$ at $x=-\infty$ and $c$ at
$x=+\infty$. Irrespective of integrability, the conservation of
topological charge implies that this sector is preserved
under time evolution. But if the model is integrable, then the sector
can be further restricted to the (possibly empty) subspace 
of two-kink states with masses $M_i\neq M_j$, 
and rapidities $\theta_i\neq \theta_j$. If we 
take $\theta_i>\theta_j$, then for the {\it
in} states, found as $t\rightarrow-\infty$, the kink of mass $M_i$ lies
to the left of the kink of mass $M_j$. Hence $D_{in}$, the dimension
of the space of {\it in} states, is equal to $\sum_b
I^{(i)}_{ab}I^{(j)}_{bc}$. Now since $M_i\neq M_j$ and the model is
integrable, reflection is ruled out and the space of {\it out} states
is spanned by states in which the kink of mass $M_i$ is to the right
of the kink of mass $M_j$. Thus, $D_{out}=\sum_b
I^{(j)}_{ab}I^{(i)}_{bc}$. But since we can map from one space to the
other and back by a unitary matrix $S$ and its inverse, the two
dimensions must be equal. That is, $D_{in}=D_{out}$, which when taken
for all pairs of vacua $a,b$ is exactly the statement
$[I^{(i)},I^{(j)}]=0$. Note that the argument holds even when
some kinks have multiplicities higher than one, and also when
there are some kinks of zero topological charge. Both situations will
be encountered below.

In this paper attention is being restricted to the ground-state energy.
Even so, the formula~(\ref{Easympt}) turns out to contain a large
amount of information about the kink structure. To extract the
asymptotics from the massive TBA equations, an iterative approach can
be used~\cite{Zc}. To first approximation the massive pseudoenergies
behave as
$\ep^{\sscr(1)}_i(\theta)\sim\nu^{\sscr(1)}_i(\theta)=M_iR\cosh\theta$, 
while the 
others take the constant values $\log\CX_i^{(\alpha)},$
as given in appendix~C. After one iteration,
\eq
\ep_i^{(1)}(\theta)=M_iR\cosh(\theta)+\sum_{j,\beta}
N_{ij}^{1\beta}\log\lf(1+{\CX_j^{(\beta)}}^{-1}\ri)~,
\label{epcorr}
\en
where
\[
N_{ij}^{\alpha\beta}=\frac{1}{2\pi}\iint\psi^{\alpha\beta}_{ij}(\theta)
\,d\theta
\]
with $\psi^{\alpha\beta}_{ij}$ a generic notation for the kernel
linking $\ep^{(\alpha)}_i$ to $\ep^{(\beta)}_j$ in the TBA system.
Hence
\bea
E(R)&=& -\frac{1}{2\pi}\sum_i\iint\!d\theta\,
 M_i\cosh\theta\log\lf[1+e^{-\ep^{(1)}_i (\theta)}\ri]\nn\\
&\sim & -\frac{1}{2\pi}\sum_i
\iint\!d\theta\, M_i\cosh\theta\, e^{-M_iR\cosh\theta}\prod_{j,\beta}
\lf(1+{\CX_j^{(\beta)}}^{-1}\ri)^{N_{ij}^{1\beta}}~. \nn \\ \nn
\eea
Identifying the factors $\Lambda_i(R)=
\frac{1}{2\pi}\iint\,d\theta M_i\cosh\theta e^{-M_iR\cosh\theta}$, the
eigenvalues $\lambda_0^{(i)}$ can be read off:
\eq
\lambda_0^{(i)}= \prod_{j,\beta}
\lf(1+{\CX_j^{(\beta)}}^{-1}\ri)^{N_{ij}^{1\beta}}~.
\en
For the calculations it is actually easier to obtain the correction
term in~(\ref{epcorr}) by substitution into the relevant Y-system.
The final results are remarkably simple: there are $n$ different kink
masses $M_i$, $i=1\dots n$, equal to the corresponding sine-Gordon
masses (\ref{bmasses}), (\ref{bbmasses}), and for the ground state
energy we have
\bea
\lambda^{(i)}_0&=&i{+}1\qquad (i=1\dots n{-}1)\,,\nn\\
\lambda^{(n)}_0&=&\sqrt{N}\,.\\ \nn
\eea
\vskip -10pt

It remains to identify a set of commuting, non-negative 
integer valued, $(N{+}1){\times}(N{+}1)$,
symmetric and $\ZZ_N$-symmetric matrices $I^{(1)}\dots I^{(n)}$ for
which these form the largest (in the sense of 
equation~(\ref{lambdaorder})\,) set of simultaneous eigenvalues. 
In fact, the eigenvalues are individually maximal.
To see this, first recombine the incidence matrices as
$I_{ab}(R)=\sum_i\Lambda_i(R)I^{(i)}_{ab}$. For $R$ large enough, the
eigenvector of $I_{ab}(R)$ with the largest eigenvalue is also the
simultaneous eigenvector that we are looking for.
Furthermore, for $R$ finite $I_{ab}(R)$ is connected
(note, the same cannot be assumed for the individual $I^{(i)}_{ab}$).
Hence, we can apply the Perron-Frobenius theorem to deduce that this
eigenvector is $\ZN$-symmetric, with all components of like sign.
Using the theorem in reverse, this eigenvector must be maximal for the
individual incidence matrices.
Two of these matrices can be
written down almost immediately. Up to a relabelling of the nodes 
$1\dots N$ corresponding to the ordered vacua, they are:
\eq
I^{(1)}= 
\left( 
       \begin{tabular}{c|c}
                      &  0 \\
{\Large $[\widehat{A}_{N-1}]_{ab}$ }&  . \\
                      &  . \\ \hline
     \rll ~  0~~~.~~~.~~~  & ~ 2
       \end{tabular}
\right)
\quad,\quad
I^{(n)}=
\left( 
       \begin{tabular}{c|c}
                      &  1 \\
         {\Large $0$ }&  1 \\
                      &  . \\ \hline
      \rll ~ 1~~~1~~~.~~~  & ~ 0
       \end{tabular}
\right)~~
\label{Ioneandn}
\en
where $\widehat A_{N{-}1}$ is the $N{\times}N$ incidence matrix of the affine
$a^{(1)}_{N{-}1}$ Dynkin diagram.
So long as $N$ is not a perfect square, $I^{(n)}$ is fixed uniquely,
while for $I^{(1)}$ we have to add a
connectivity assumption when $N$ is not prime.  In fact, any of the
other possibilities for $I^{(1)}$ would have 
the kinks of mass $M_1$ joining
non-adjacent ordered vacua. At least
in a classical picture of the kinks, this can be ruled out since $M_1$ is
minimal among the remaining kink masses.
There only remains the possibility for
one of the blocks making up $I^{(1)}$ to be equal to zero, and
imposing $[I^{(1)},I^{(n)}]=0$ rules this out as well.

For the remaining kinks it is not possible to be so definite on the
basis of the ground-state energy alone. 
However, one set of conjectures seems natural. Note first that the forms of
$I^{(1)}$ and $I^{(n)}$ are consistent with the idea that the kinks of 
mass $M_1$ are the (lowest-mass) bound states of a set of fundamental kinks 
with mass $M_n{=}M$. This parallels the related sine-Gordon 
model, where $M_n$ is the mass of the soliton, and $M_1$ the mass of its 
first bound state -- the first breather. 
If the analogy is to extend, then the kinks of mass $M_j$ in the
$\ZZ_N$ model should be bound states of $j$ of the kinks of mass $M_1$.
The allowed topological charges should therefore be found among the paths 
with $j$ steps on the graph of $I^{(1)}$. Adding this to the other 
requirements leads to the proposal
\eq
I^{(j)}=
\left( 
       \begin{tabular}{c|c}
                      &  0 \\
{\large min}{\Large 
$\lf([\widehat{A}^{~j}_{N-1}]_{ab}\,,\hbox{\large $1$}\ri)$} &  . \\
                      &  . \\ \hline
      \rll  0~~~.~~~.~~~.~~~  & ~ j+1
       \end{tabular}
\right)\qquad (j=1\dots n{-}1)\,.
\label{Ibreathers}
\en
Commutativity with $I^{(n)}$ was used to
settle the value of $I^{(j)}_{N{+}1,N{+}1}\,$. An alternative 
characterisation is the statement 
that the breather-related incidence matrices
form part of an $SU(2)$-type fusion hierarchy:
defining $I^{(0)}=\II$, the 
identity matrix, 
and $I^{(1)}$ as in equation~(\ref{Ioneandn}), the matrices
$I^{(2)}\dots I^{(n{-}1)}$ are determined by the relation~(\ref{fuse}),
just as for the $\ep^{(1)}$ perturbation, though with one more vacuum,
and different initial conditions.
Note that if $I^{(1)}$ and $I^{(n)}$ commute and have the correct
Perron-Frobenius eigenvalues, then the same automatically holds for
$I^{(2)}\dots I^{(n-1)}$, given the fusion property.
Some preliminary investigations of excited-state spectra for the 
$b_n$-related cases are also consistent with~(\ref{Ibreathers}).

\smallskip
\begin{figure}[htb]
\hspace{3.0cm}
\epsfxsize=200pt
\epsfysize=80pt
\epsfbox{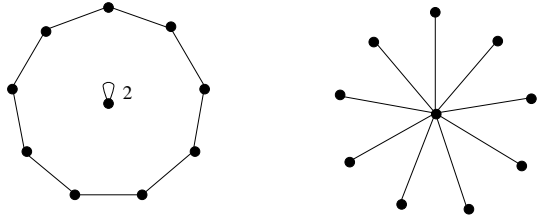}
\caption{The matrices $I^{(1)}$ and $I^{(2)}$ for
the massive $\ZZ_9$ theory }
\label{z9pic} 
\end{figure}
\begin{figure}[htb]
\hspace{1.0cm}
\epsfxsize=320pt
\epsfysize=95pt
 \epsfbox{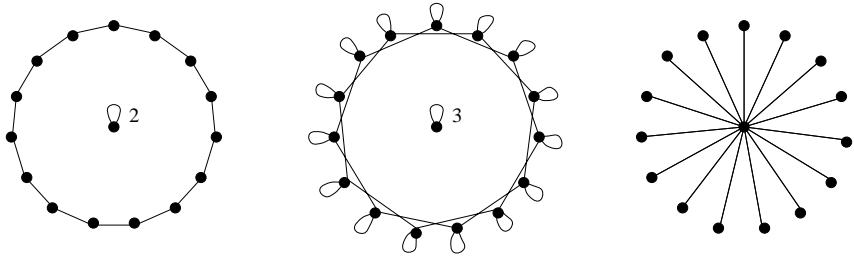}
\caption{The matrices $I^{(1)}$, $I^{(2)}$ and $I^{(3)}$ for
the massive $\ZZ_{15}$ theory }
 \label{z15pic} 
\end{figure}

Figures~\ref{z9pic} and~\ref{z15pic} illustrate the general
idea with the kink structures for $N{=}9$ and $N{=}15$. Tunneling to and 
from the disordered phase only occurs via the fundamental-kink 
instantons, while instantons associated with an increasingly intricate 
structure of higher kinks serve to connect the various ordered phases. 
Note also the presence of tadpoles in all but the fundamental set of kinks.
These have been seen before~\cite{Zh}, and their appearance in the
finite-size effects can be understood either 
semi-classically, as corrections due to
instantons of zero topological charge, or else as being due
to virtual particle-like excitations (`breathers') above 
the various 
vacua.  Either way, the predicted effect on the large-$R$ asymptotics is the 
same, and has indeed been observed in some other situations~\cite{RSTa}.
A more novel feature in this context is that, for $N{>}16$,
some kinks appear in more than one incidence matrix from
the set $I^{(1)}\dots I^{(n)}$.
In other words, some pairs of vacua are joined not only by a simple kink of
minimal mass, but also by excitations of this kink with
higher masses -- these can be thought of as  breathers with non-zero
topological charge.

To see this aspect most clearly, the spectrum of asymptotic
one-particle states can be reorganised according to their topological
charges. Let $a,a{+}i$ always label ordered vacua $1\dots N$, 
with $N{+}1$ the disordered vacuum. With $n$ as ever equal to the
integer part of $(N{-}1)/4$, the prediction is that the massive flow should
exhibit:
{\begin{list}{$\bullet$}
{\setlength{\leftmargin}{12pt}\setlength{\parsep}{1pt}}
\item for $i=1,2,\dots n{-}1\,$: a kink joining each
$a$ to $a{+}i$ with
mass $M_i\,$, and excitations of this kink with masses $M_j\,$,
$j=i{+}2,i{+}4,\dots\le n{-}1\,$;
\item kinks joining $N{+}1$ to each $a$ with mass $M_n\,$;
\item excitations of each ordered vacuum with masses $M_j$, 
$j=2,4,\dots\le n{-}1\,$;
\item excitations of the disordered vacuum with masses
$M_1,M_2,\dots M_{n{-}1}\,$.
\end{list}}
\vskip -3pt
\noindent
All appear with unit multiplicity at each mass
except for those in the last line, which
have multiplicities $2,3,\dots n$.

\medskip
\noindent
The discussion in this subsection has been based on a
physical input to determine how the $\ZZ_N$ symmetry acts on the vacua: 
we required one singlet and one $N$-element multiplet.
Whenever $N$ factorises, say as $N{=}PQ$, then it is possible to use an
orbifold construction~\cite{FGa} to produce alternative patterns of vacua, 
which might also be described by the self-dual $\ZZ_N$ TBA systems. 
The incidence matrices now split into blocks of sizes
$P$ and $Q$, rather than the previous $N$ and $1$. The basic matrices are:
\[
I^{(1)}= 
\left( 
       \begin{tabular}{c|c}
                      &   \\
{\Large $[\widehat{A}_{P-1}]_{ab}$ }&  {\Large $0$} \\
                      &   \\ \hline
     \rlll ~~~~~~{\Large $0$}~~~~~~~  & 
{}~{\Large $[\widehat{A}_{Q-1}]_{cd}$ }
       \end{tabular}
\right)
{}~~,\quad
I^{(n)}=
\left( 
       \begin{tabular}{c|c}
                      &   \\
         {\Large $0$ }&  {\Large $1_{ad}$} \\
                      &   \\ \hline
\rlll ~~~~~{\Large $1^{\mbox{\normalsize\sf t}}_{cb}$}~~~~~
                      & ~~~{\Large $0$}~~~{}
       \end{tabular}
\right)
\]
where $\widehat A_{P{-}1}$ and 
$\widehat A_{Q{-}1}$ are the incidence matrices of the affine
$a^{(1)}_{P{-}1}$ and $a^{(1)}_{Q{-}1}$ Dynkin diagrams, and ${\bf 1}_{ad}$
is a $P{\times}Q$ matrix with all entries equal to~$1$. The remaining
matrices, $I^{(2)}\dots I^{(n{-}1)}$, follow from $I^{(1)}$ via the fusion
relation~(\ref{fuse}), with $I^{(0)}=\II$. Such systems of vacua and
kinks could be relevant to first-order transition regions  where
$\ZZ_P$ and $\ZZ_Q$ ordered phases coexist.

%
\resection{Conclusions}
The checks performed in this paper have been rather exhaustive, and seem 
to us to establish beyond reasonable doubt that each pair of
the self-dual $\ZN$ TBA systems listed in section~2 does indeed
describe a pair of flows from the $\ZN$-symmetric conformal field 
theory into
massless and massive phases, with the two flows being
related by a change in the sign of a coupling constant.
Given the earlier results of ref.~\cite{JMOa},
this effectively 
settles the question posed in the introduction about the
location of the Fateev-Zamolodchikov Boltzman weights
in the larger phase diagram.  
In addition, a number of predictions have
been obtained which deserve further exploration.

One important outstanding question is to find
a collection of S-matrices, both massive and massless, lying behind
the TBA systems that have been proposed. In fact, a complete set of
S-matrices for the massive flows was suggested some years ago by 
Fateev~\cite{Fa}, but we have not been able to reconcile these with
the TBA results reported in section~4.3. In particular, Fateev
predicted that for $N$ odd there should be $(N{-}5)/2$ bound states of
the fundamental kinks, all with different masses, while the
asymptotics of the TBA systems proposed above predict only
$[(N{-}5)/4]$. If the TBA systems are correct, it is very hard to see
how about half of the kink masses could be lost from the instanton
gas calculation. Given the stringent tests to which
these systems have been 
subject in this paper, it is possible that the
resolution will be found in some modification of Fateev's proposals.

On a more mathematical note,
the Y-systems and dilogarithm sum rules found here for $N$ odd appear to
be new. Past experience (see for example ref.~\cite{NRTa}) would
suggest the existence of new fermionic representations
for the characters of the $\ZN$-symmetric conformal field theories,
and possibly also an alternative integrable
deformation of the Fateev-Zamolodchikov Boltzman weights~\cite{FZa}
which, contrary to that discussed by Jimbo {\it et al}~\cite{JMOa}, 
leaves the $\ZN$ symmetry unbroken.

Further study
of the massless flows should be rewarding: as
highlighted in section~4.2, the systems introduced in this paper add to 
an already-rich collection of flows arriving at $c{=}1$. The precise
relationship between these models needs to be elucidated, and the
possibility to extract at least some
infrared information exactly should help in this task. 
Since the various infrared limits can all be regarded
as irrelevant perturbations of a single free boson, they should also
provide the simplest possible playground to explore more general
issues in non-renormalisable field theory.

Finally, there remain many unresolved
features of the larger phase diagrams of the $\ZN$ spin systems~\cite{DMSa},
beyond the particular self-dual directions out of the Fateev-Zamolodchikov 
points that have occupied our attention in this paper. We hope that these
will
provide further opportunities to apply the ideas of perturbed conformal field
theory and the TBA to problems of independent statistical-mechanical
interest.

%
%
\vskip 0.6cm
\noindent
{\bf Acknowledgements} -- We would like to thank
John Cardy, G\"unter von Gehlen, Ferdinando Gliozzi, Francesco Ravanini
and Robert Weston for useful 
discussions. PED would also like to thank SPhT Saclay and the Benasque 
Centre for Physics for hospitality while some of this work was in
progress.  PED and KET thank the EPSRC for 
an Advanced Fellowship and a Research Studentship respectively, and
RT thanks the Mathematics Department of Durham University for a
postdoctoral fellowship.
This work was supported in part by a Human
Capital and Mobility grant of the European Union, contract number
ERBCHRXCT920069, and in part by a NATO grant, number CRG950751.

%
\appendix
\section{Sine-Gordon and kernel data}
\setcounter{equation}{0}
The sine-Gordon model can be defined by the Lagrangian
\eq
{\cal L}={1\over 2}(\de_{\mu}\varphi)^2+{m^2\over\beta^2}\cos\beta\varphi
\label{SGact}
\en
and results in a family of scattering theories, conveniently
parametrised by either $p$ or $h$, where
\eq
p={2\over h}={\beta^2\over 8\pi-\beta^2}~.
\label{phdef}
\en
It will also be useful to define $N=2h{+}4=32\pi/\beta^2$, 
and to set $n$ equal to the integer part of $(N{-}1)/4$.
The spectrum contains a soliton-antisoliton doublet $(s,\bar s)$ and,
for $h>2$, $n{-}1$ soliton-antisoliton bound states, or
breathers. If $s$ and $\bar s$ have mass $M$, then the breather
masses are 
\eq
M_k=2M\sin{\pi k\over h}=2M\sin{2\pi k\over N{-}4}
\quad,\quad k=1,2,\dots n{-}1~.
\label{bmasses}
\en
The soliton labels $s$ and $\bar s$ will 
often be replaced by $n$ and $n{+}1$, 
and accordingly 
\eq
M_n=M_{n{+}1}=M~.
\label{bbmasses}
\en
Scattering among solitons and antisolitons is generally
non-diagonal; the amplitudes can be found in~\cite{ZZa}. 
However, once a breather is involved the scattering becomes
diagonal. By analogy with a notation used for the affine Toda
theories, define the blocks
\[
\usbl{x}(\theta)={\sinh\bigl({\theta\over 2}+{\imath\pi x\over 2h}\bigr)\over
        \sinh\bigl({\theta\over 2}-{\imath\pi x\over 2h}\bigr)}\quad;
\quad\ubl{x}=\usbl{x-1}\usbl{x+1}~.
\]
Then 
\eq
S_{jk}=
 \prod^{ \atop j+k-1}_{|j-k|+1\atop {\rm
step~2}}\ubl{l}\ubl{h-l}
\qquad (j,k=1\dots n{-}1)~,
\label{Sjk}
\en
and
\eq
S_{kn}=S_{k,n{+}1}=
(-1)^k\prod^{ \atop h/2+k-1}_{h/2-k+1 \atop {\rm step~2}}\ubl{l}
\qquad (k=1\dots n{-}1)~.
\label{Sks}
\en

The couplings relevant for the $b_n$ and $d_{n{+}1}$ series of TBA systems
correspond to $h=2n{-}1$ and $h=2n$ respectively -- the dual Coxeter
numbers of $b_n$ and $d_{n{+}1}$. When $h=2n$, the entire S-matrix is diagonal
with the remaining S-matrix elements given by
\bea
\llap{n~{\rm even}}~:~~S_{nn}=S_{n{+}1,n{+}1}=
\prod^{\atop 2n-3}_{l=1\atop {\rm step~4}}\ubl{l}&;& 
S_{n,n{+}1}=\prod^{\atop 2n-1}_{l=3\atop {\rm step~4}}\ubl{l}~;\nn\\
\llap{n~{\rm odd}}~:~~S_{nn}=S_{n{+}1,n{+}1}=
\prod^{\atop 2n-1}_{l=1\atop {\rm step~4}}\ubl{l}&;& 
S_{n,n{+}1}=-\!\!\prod^{\atop 2n-3}_{l=3\atop {\rm step~4}}\ubl{l}~.
\label{Sss} \\ \nn
\eea
\vskip -5pt\noindent
In this case, the various prefactors of $-1$ (which anyway make no
difference to the definitions below) 
can be omitted to leave the minimal version 
of the $d_{n+1}$ Toda S-matrix.

The TBA systems in the text involve certain kernels. The
first of these are related to the
logarithmic derivatives of the diagonal sine-Gordon S-matrix elements:
\eq
\phi_{jk}=-\imath\ddt\log S_{jk}\quad;\quad
\psi_{jk}=-\imath\ddt\log T_{jk}\,.
\label{kernels}
\en
Here at least one of the indices $j$ and $k$ must have a
`breather' value,  $1,2,\dots n{-}1$.
In the definition of $\psi_{jk}$, the function $T_{jk}$ is 
obtained by replacing each
block $\ubl{x}$ in (\ref{Sjk}--\ref{Sks}) by $\usbl{x}$. The kernels
$\psi_{jk}$ do not appear in the TBA equations for the sine-Gordon 
model itself, but rather they are part of the conjectures
that are the main topic of this paper. Their forms 
are a natural generalisation of previously-known examples~\cite{Ra}.

For the
$d_{n+1}$-related models the definition can be extended immediately to
cover the remaining cases when both $j$ and $k$ take the values 
$n$ or $n{+}1$.
Otherwise,
the non-diagonal scattering of the solitons means that the
associated kernels are more elaborate. Define
an integer~$\rho$ by $N=4n{+}\rho$, and then set
\eq
\chi_{\rho}(\theta)=\frac{2h}{\rho\cosh{\fract{2h}{\rho}\theta}}~~.
\en
This function has the property that
\[
\chi_{\rho}{*}f(\theta{+}\fract{\imath\pi\rho}{4h})
+\chi_{\rho}{*}f(\theta{-}\fract{\imath\pi\rho}{4h})=f(\theta)~.
\]
The soliton-soliton S-matrix consists of a `scalar' piece multiplying
a non-diagonal matrix. Associated with the first factor is a
pseudoenergy $\ep_n(\theta)\,$,
with interaction kernels in the TBA given by
\eq
\phi_{nn}(\theta)=\chi_{\rho}{*}\phi_{n,n{-}1}(\theta)\quad,\quad
\psi_{nn}(\theta)=\chi_{\rho}{*}\psi_{n,n{-}1}(\theta)~.
\en
(For $N{\le}7$, $\phi_{n,n{-}1}$ and $\psi_{n,n{-}1}$ are not defined
and we set $\phi_{nn}{=}\psi_{nn}{=}0$.)
In addition, depending on the continued-fraction expansion of $2/h$, 
the non-diagonal part induces a
number of fictitious `magnonic' particles~\cite{TSa}, 
together with their attendant kernels.
We will only need those which arise when $N$ is an
integer; these can be obtained from the Y-systems of ref.~\cite{Tb},
and are
\eq
\phi_1(\theta)={h\over\cosh h\theta}
\nn\en
\bea\ds
\phi_2(\theta)={2h\over \cosh 2h\theta}\,&&\qquad
\phi_4(\theta)={8h\cosh{2h\over 3}\theta\over 
 3(4\cosh^2{2h\over 3}\theta-3)}\nn\\
\phi_3(\theta)={{2\over 3}h\over \cosh {2h\over 3}\theta}&&\qquad
\phi_5(\theta)={8h\cosh{2h\over 3}\theta\over 
\sqrt{3}(4\cosh^2{2h\over 3}\theta-1)}\,~.
\label{magkernels}\\
\nn
\eea
As above, $h$ is equal to $N/2-2\,$.
%
%
%
%
%
%
%
%
\section{Y-systems}
\setcounter{equation}{0}

This section records the self-dual $\ZZ_N$ 
Y-systems. Their derivation from the
TBA equations is rather involved, but follows the same lines as the
simpler cases explained in ref.~\cite{RTVa}. In particular, the energy
terms are eliminated using the following variant of the
Perron-Frobenius mass property of the $ade$-related theories:
\bea
2\cos(\fract{\pi}{h})\,M_i &=&  \sum_j l^{[a_{n{-}1}]}_{ij} M_j \quad
(i=1,\dots n{-}2)\,; \nn \\
2\cos(\fract{\pi}{h})\,M_{n-1}&=&M_{n-2}+2\cos(\fract{(4{-}\rho)\pi}{4h})
M_{n}\,;\nn \\[2pt]
2\cos(\fract{\pi\rho}{4h})\,M_{n} &=&  M_{n-1}\,, \nn
\eea
where $N=4n{+}\rho$, as in the last section, and $l^{[a_{n{-}1}]}_{ij}$ 
is the incidence matrix of the $a_{n{-}1}$ Dynkin diagram. 

\begin{figure}[p] 
\hspace{1.0cm}
\epsfxsize=350pt 
\epsfysize=400pt
\epsfbox{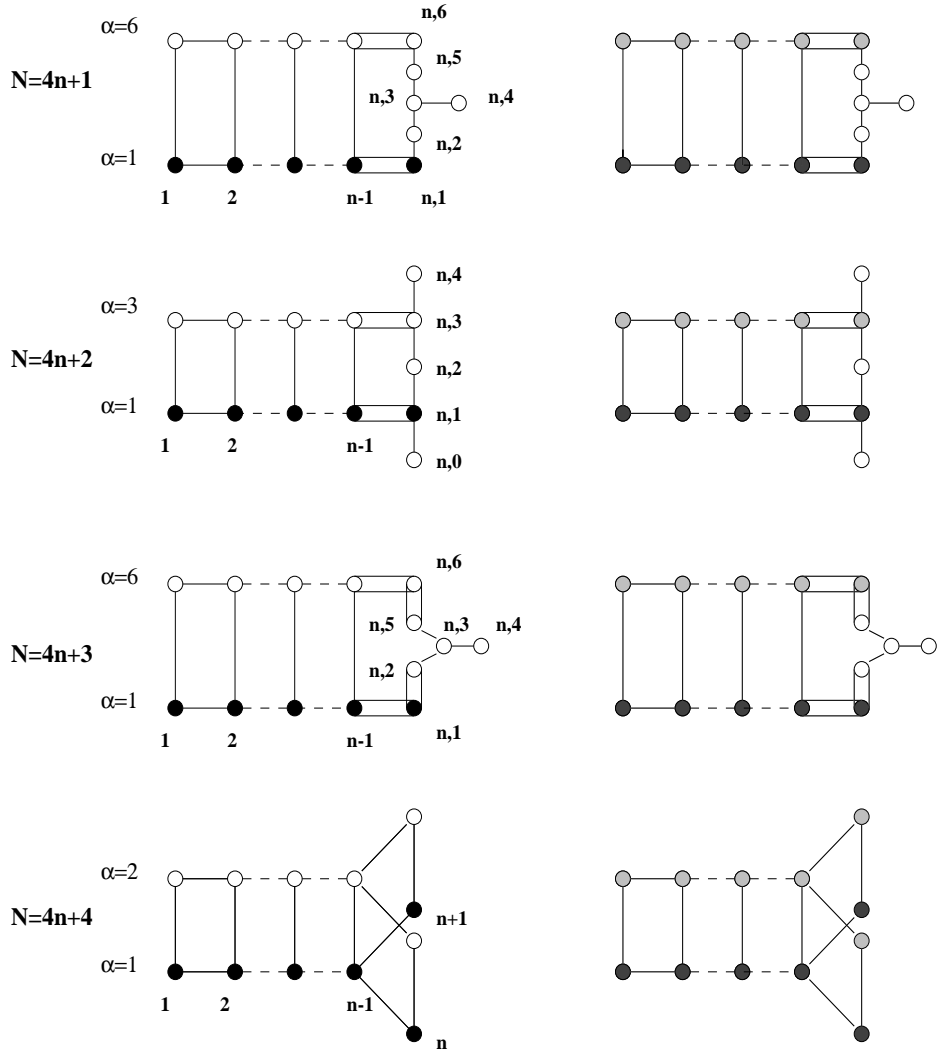} 
\caption{Diagrammatic representations of the $\ZZ_N$ Y-systems. 
The left/right columns represent the 
massive/massless theories respectively.}\label{Z2Y} 
\end{figure}

The results are
$\ZZ_{2}$-symmetrised versions of the sine-Gordon Y-systems
constructed in ref.~\cite{Tb}. 
Figure~\ref{Z2Y} shows a diagrammatic representation,
following refs.~\cite{Ra,Tb}.
In order to write them down in a managable form, define
 \begin{eqnarray}
\overline{Y}_{i}^{(\alpha)}[\theta,r]
&=&  Y_{i}^{(\alpha)}(\theta -\fract{\imath\pi r}{h})
\,Y_{i}^{(\alpha)}(\theta +\fract{\imath\pi r}{h})\nn\\
\overline{Y}_{i}^{(\alpha)}\{\theta, r\}
&=& \Big( 1 +
Y_{i}^{(\alpha)}(\theta-\fract{\imath\pi r}{h})\Big)
\Big( 1 + Y_{i}^{(\alpha)}(\theta +\fract{\imath\pi r}{h})
\Big)\nn\\
V_{i}^{(\alpha)}\{\theta, r\} &=&
\Big(1 + Y_{i}^{(\alpha)}(\theta
-\fract{\imath\pi r}{h})^{-1}\Big)^{-1}
\Big(1 + Y_{i}^{(\alpha)}(\theta
+\fract{\imath\pi r}{h})^{-1}\Big)^{-1}\nn
 \end{eqnarray}
With this notation in place, the Y-systems for the four
cases can be written as follows.
%
\\[0.25cm]
{\bf{1) $N=4n{+}1$}}
\bea
\noalign{\noindent Nodes $(i{<}n{-}1,\alpha=1,6)$, with
$\wtilde\alpha=7{-}\alpha\,:$\vs}
\overline{Y}_{i}^{(\alpha)}[\theta, 1]&=&
\Big(1+ Y_{i}^{(\wtilde\alpha)}(\theta)^{-1}\Big)^{-1}
 \prod_{j=1}^{n-1} \Big(1 + Y_{j}^{(\alpha)}(\theta)
\Big)^{l^{[a_{n-1}]}_{ij}} \label{common}\\[4pt]
\noalign{\noindent Nodes $(n{-}1,\alpha{=}1,6)~$:\vs}
\overline{Y}_{n-1}^{(1)}[\theta, 1]&=&
\Big( 1 + Y_{n-1}^{(6)}(\theta)^{-1}\Big)^{-1}
\Big(1+Y_{n-2}^{(1)}(\theta)\Big)
\Big(1+Y_{n}^{(4)}(\theta)\Big)
\Big(1+Y_{n}^{(5)}(\theta)\Big) \nn\\
&&\quad~~\times \,
\overline{Y}_{n}^{(1)} \{\theta,3/4\}
 \overline{Y}_{n}^{(2)} \{\theta,2/4\}
 \overline{Y}_{n}^{(3)} \{\theta,1/4\}\nn\\[8pt]
\overline{Y}_{n-1}^{(6)}[\theta, 1]&=&
\Big( 1 + Y_{n-1}^{(1)}(\theta)^{-1}\Big)^{-1}
\Big(1+Y_{n-2}^{(6)}(\theta)\Big) 
\Big(1+Y_{n}^{(4)}(\theta)\Big)
\Big(1+Y_{n}^{(2)}(\theta)\Big) \nn\\
&&\quad~~\times \,
\overline{Y}_{n}^{(6)} \{\theta,3/4\}
 \overline{Y}_{n}^{(5)} \{\theta,2/4\}
 \overline{Y}_{n}^{(3)} \{\theta,1/4\}\nn\\[4pt]
\noalign{\noindent Nodes $(n,\alpha{=}1\dots 6)~$:\vs}
\overline{Y}_{n}^{(\alpha)}[\theta, 1/4] &=&
\Big(1+Y_{n-1}^{(\alpha)}(\theta)\Big) \prod_{\beta}
\Big(1+ Y_{n}^{(\beta)(\theta)^{-1}}
\Big)^{-l^{[e_{6}]}_{\alpha \beta}}\nn
\eea
with $Y_{n-1}^{(\alpha)}(\theta) \equiv 0$ when $\alpha
\neq 1,6$.
\\[0.5cm]
%
\goodbreak
\noindent
{\bf{2) $N=4n{+}2$ ($b_{n}$) }}\nobreak
\bea
\noalign{\noindent
Nodes $(i{<}n{-}1,\alpha{=}1,3)\,$: equation~(\ref{common})
holds with $\wtilde\alpha=4{-}\alpha\,$.\vs}
\noalign{\noindent
Nodes $(n{-}1,\alpha{=}1,3)\,$:\vs}
\overline{Y}_{n-1}^{(\alpha)}[\theta, 1]&=&
\Big( 1 + Y_{n-1}^{(\wtilde\alpha)}(\theta)^{-1}\Big)^{-1}
\Big(1+Y_{n-2}^{(\alpha)}(\theta)\Big)
\Big(1 + Y_{n}^{(\alpha-1)}(\theta)\Big)
\nn\\
&&\quad~~\times\,
\Big(1+Y_{n}^{(\alpha+1)}(\theta)\Big)
\overline{Y}_{n}^{(\alpha)} \{\theta,1/2\} 
\nn\\
\noalign{\noindent
Nodes $(n,\alpha{=}0\dots 4)\,$
(with $Y_{n-1}^{(\alpha)}(\theta) \equiv 0$ when
$\alpha \neq 1,3$)~:\vs}
\overline{Y}_{n}^{(\alpha)}[\theta, 1/2]&=&\Big(1 +
Y_{n-1}^{(\alpha)}(\theta)\Big)
\prod_{\beta} \Big(1+
Y_{n}^{(\beta) }(\theta)^{-1}\Big)^{-l^{[a_{5}]}_{\alpha \beta}}
\nn
\eea
 \\[0.2cm]
%
%
{\bf{3) $N=4n{+}3$}}
\bea
\noalign{\noindent
Nodes $(i{<}n{-}1,\alpha{=}1,3)\,$: equation~(\ref{common})
holds with $\wtilde\alpha=7{-}\alpha\,$.\vs}
\noalign{\noindent
Nodes $(n{-}1,\alpha{=}1,6)\,$:\vs}
\overline{Y}_{n-1}^{(1)}[\theta, 1]&=&\Big(1+Y_{n-2}^{(1)}(\theta)\Big)
\Big(1+Y_{n}^{(2)}(\theta)\Big) 
\Big( 1+Y_{n-1}^{(6)}(\theta)^{-1}\Big)^{-1}
\,\overline{Y}_{n}^{(1)} \{\theta,1/4\} \nn\\
\overline{Y}_{n-1}^{(6)}[\theta, 1]&=&\Big(1+Y_{n-2}^{(6)}(\theta)\Big)
\Big(1+Y_{n}^{(5)}(\theta)\Big) 
\Big( 1 +Y_{n-1}^{(1)}(\theta)^{-1}\Big)^{-1}
\,\overline{Y}_{n}^{(6)} \{\theta,1/4\} \nn\\[4pt]
\noalign{\noindent
Nodes $(n,\alpha=1\dots 6)\,$:\vs}
\overline{Y}_{n}^{(1)}[\theta,3/4]&=&
\Big(1+Y_{n-1}^{(1)}(\theta)\Big) \Big( 1
+Y_{n}^{(4)}(\theta)^{-1}\Big)^{-1}
\Big( 1 +Y_{n}^{(5)}(\theta)^{-1}\Big)^{-1}\nn\\
&&\quad~~\times\, V_{n}^{(2)}\{\theta,2/4\}\,V_{n}^{(3)}\{\theta,1/4\}\nn\\
\overline{Y}_{n}^{(2)}[\theta,1/4]&=&\Big(1+Y_{n}^{(3)}(\theta)\Big)
\Big( 1 +Y_{n}^{(1)}(\theta)^{-1}\Big)^{-1} \nn\\
\overline{Y}_{n}^{(3)}[\theta,1/4]&=&
\Big(1+Y_{n}^{(4)}(\theta)\Big)\Big(1+Y_{n}^{(2)}(\theta)\Big)
\Big(1+Y_{n}^{(5)}(\theta)\Big)\nn\\
\overline{Y}_{n}^{(4)}[\theta,1/4]&=&\Big(1+Y_{n}^{(3)}(\theta)\Big)\nn\\
\overline{Y}_{n}^{(5)}[\theta,1/4]&=& \Big(1+Y_{n}^{(3)}(\theta)\Big)
\Big( 1 +Y_{n}^{(6)}(\theta)^{-1}\Big)^{-1}\nn\\
\overline{Y}_{n}^{(6)}[\theta,3/4]&=&
\Big(1+Y_{n-1}^{(6)}(\theta)\Big) \Big( 1
+Y_{n}^{(4)}(\theta)^{-1}\Big)^{-1}
\Big( 1 +Y_{n}^{(2)}(\theta)^{-1}\Big)^{-1}\nn\\
&&\quad~~\times\, V_{n}^{(5)}\{\theta,2/4\}\,V_{n}^{(3)}\{\theta,1/4\}
\nn
\eea
%
{\bf{4) $ N = 4n{+}4$ ($d_{n+1}$)}}
\bea
\noalign{\noindent
In this case the scope of equation~(\ref{common}),
with $\wtilde\alpha=3{-}\alpha$, 
can be extended to cover all of the nodes 
$(i{=}1\dots n{+}1,\alpha{=}1,2)\,$:\vs}
\overline{Y}_{i}^{(\alpha)}[\theta, 1]&=&
\Big(1+ Y_{i}^{(\wtilde\alpha)}(\theta)^{-1} \Big)^{-1}
 \prod_{j=1}^{n+1} \Big(1 + Y_{j}^{(\alpha)}(\theta)
\Big)^{l^{[d_{n{+}1}]}_{ij}}
\nn
\eea
\bigskip\bigskip

\noindent
In the above formulae, $l^{[a_{n-1}]}_{ij}$,
$l^{[d_{n+1}]}_{ij}$,
$l^{[a_5]}_{\alpha\beta}$ and
$l^{[e_6]}_{\alpha\beta}$  are the incidence matrices of the
corresponding Dynkin diagrams.

\medskip\noindent
We checked numerically the periodicity implied by these Y-systems up
to $N{=}30$, and found
\eq
Y_i^{(\alpha)} \lf( \theta + \imath  \pi { N+2 \over N-4}  \ri)=
Y_i^{(\wtilde\alpha)} \lf( \theta  \ri)~,
\en
which in turn implies the result~(\ref{yfulper}) quoted in the main
text.

%
%
\section{Dilogarithm sum rules}
\setcounter{equation}{0}
The sum rules are conveniently given in terms of certain limits of the
functions $Y^{(\alpha)}_i(\theta)=e^{\ep_i^{(\alpha)}(\theta)}$. With
the $\ep^{(\alpha)}_i(\theta)$ the solution of a {\it massless} system,
define
\bea
\Y_{i}^{(\alpha)} &=& \lim_{\MR\rightarrow 0} Y_i^{(\alpha)}(\theta)
\sps \sps ( \theta~\mbox{finite})
\virg\nn\\
\CX_{i}^{(\alpha)} &=& \lim_{\theta \rightarrow\infty }
Y_i^{(\alpha)}(\theta)
\sps \sps (\MR ~\mbox{finite})
\virg\nn\\ 
\CZ_{i}^{(\alpha)}&=&\lim_{\MR\rightarrow\infty } Y_i^{(\alpha)}(\theta)
\sps \sps (\theta~\mbox{finite})~.
\label{ydefs}\\ \nn
\eea
\vskip -5pt\noindent
(For the pseudoenergies which solve the massive systems, the
$\Y^{(\alpha)}_i$ and $\CX^{(\alpha)}_i$ are unchanged while
$\CZ^{(\alpha)}_i=\CX^{(\alpha)}_i\,.$) The numbers defined
by~(\ref{ydefs}) furnish stationary solutions to the Y-systems of
appendix~B, subject to the following constraints: all 
$\Y^{(\alpha)}_i$ are finite;
$\CX^{(1)}_i=\infty~\forall i$ with the rest finite; and
$\CZ^{(1)}_i=\CZ^{(\wtilde 1)}_i=\infty~\forall i$ with the rest finite.
The infinite quantities do not contribute to the sum rules. To find
the values of the others, start with the following ansatz for the
$a_{n{-}1}\times a_2$ tail:
\[
\Y_i^{(\alpha)}= { \sin( (i+3) \eta)  \sin( i \eta) \over \sin( ( 2
\eta))
\sin( \eta) } \virg
\]
\[
\CX_i^{(\tilde 1)}= (i+2) i  \virg
\]
with $i=1\dots n{-}1$.
The value of $\eta$ can be found by examining the
final nodes $(n,\alpha)$; 
a positive solution results on setting  $\eta= {\pi \over N+2}$. This
also gives values
in good agreement with the direct  numerical solution of the TBA
equations.
For the remaining nodes the expressions are more complicated, and 
will be given case-by-case.

\bigskip
\noindent
{\bf 1) $N= 4n{+}1$} \\
\[
\ds{ 
\Y_{n}^{(1)}=\Y_{n}^{(6)}= { \sin^2( n \eta)   \over
\sin( (2 n+2)  \eta) \sin( 2  \eta)
}
{}~~,~~
\Y_{n}^{(2)} =\Y_{n}^{(5)} = { (\Y_{n}^{(1)})^2 \over
1+\Y_{n-1}^{(1)}-(\Y_{n}^{(1)})^2}
}
\]
\[ 
\ba{c}
\ds{
\Y_{n}^{(4)}= {   (\Y_{n}^{(1)})^{3/2}
(1+\Y_{n}^{(1)})^{1/2} \over (1 + \Y_{n-1}^{(1)} -
(\Y_{n}^{(1)})^2 ) }
} \acc
\ds{ \hspace{-0.5cm}
 \Y_{n}^{(3)}=  { (\Y_{n}^{(1)})^3 (1+\Y_{n}^{(1)}) \over 1
+ 2 \Y_{n-1}^{(1)} + (\Y_{n-1}^{(1)})^2 -
  2 (\Y_{n}^{(1)})^2 - 2 \Y_{n-1}^{(1)} (\Y_{n}^{(1)})^2 -
(\Y_{n}^{(1)})^3}
}\nn
\ea
\]
\[
\ba{c}
\ds{ \CX_{n}^{(6)}={n^2 \over 2 n+1} \virg ~~
\CX_{n}^{(5)}={n^2 \over (3n+1)(n+1)} } \acc
\ds{ \CX_{n}^{(4)}=\CX_{n}^{(2)}={n \over 3n+1}
\virg ~~
\CX_{n}^{(3)}={n^2 \over 8n^2+1+6n} }
\ea
\]
\[
\CZ_{n}^{(4)}=\CZ_{n}^{(2)}=\CZ_{n}^{(5)}={1 \over 3}
\virg ~~
\CZ_{n}^{(3)}={1 \over 8} ~.
\]
 
\noindent
{\bf 2) $N=4n{+}2$} \\
\[
\ba{c}
\ds{
\Y_{n}^{(1)}=\Y_{n}^{(3)}= { \sin^2( n \eta)   \over
\sin( (2n+2)  \eta) \sin( 2  \eta)}
\virg ~~
\Y_{n}^{(0)}=\Y_{n}^{(4)}= { \sin( n \eta)   \over \sin( (n+2)  \eta)
}
}\acc
\ds{
\Y_{n}^{(2)}= { \sin^2( n \eta)   \over \sin^2( (n+2)  \eta)
}
}
\ea
\]
\[
\ba{c}
\ds{
\CX_{n}^{(3)}= { n^2   \over 2n+1 }
~~,~~
\CX_{n}^{(2)}=\CX_{n}^{(4)}= { n   \over n+1}
~~,~~
\CX_{n}^{(0)}= { 1 }
}
\ea
\]
\[
\CZ_{n}^{(0)}=
\CZ_{n}^{(2)}=
\CZ_{n}^{(4)}=1 ~.
\]

\noindent
{\bf 3) $N=4n{+}3$} \\
\[
\ba{c}
\ds{
\Y_{n}^{(1)}=\Y_{n}^{(6)}= { \sin^2( n \eta)   \over
\sin( (2n+2)  \eta) \sin( 2  \eta)
}
}\acc
\ds{
\Y_{n}^{(4)}= { \sin( (n+2) \eta) \sin((n+1) \eta) +
\sin^2(( 2n +2) \eta) \over  \sin^2( (n+2)  \eta)
}
} \acc
\ds{
\Y_{n}^{(3)}=(\Y_{n}^{(4)})^2-1
\virg ~~
\Y_{n}^{(2)}=\Y_{n}^{(5)}={ (\Y_{n}^{(4)})^2-1 \over
(1+\Y_{n}^{(4)})^{1/2}}-1
}
\ea
\]
\[
\ba{c}
\ds{
\CX_{n}^{(6)}={n^2 \over 2n+1} \virg  ~~
\CX_{n}^{(5)}={n (3n+2) \over (n+1)^2}
} \acc
\ds{
\CX_{n}^{(3)}={(4n+3)(2n+1) \over (n+1)^2}
\virg ~~
\CX_{n}^{(4)}=\CX_{n}^{(2)}={ 3n+2 \over n+1}
}
\ea
\]
\[
\CZ_{n}^{(4)}=\CZ_{n}^{(2)}=\CZ_{n}^{(5)}= 3
\virg ~~
\CZ_{n}^{(3)}= 8 ~.
\]

\noindent
{\bf 4) $N=4n{+}4$} \\
\[
\Y_s^{(\alpha)}=\Y_{\bar{s}}^{(\alpha)}= { \sin(n\eta) \over 2
\sin(
\eta) \cos((n{+}1)\eta)  }\qquad (\alpha=1,2)~,
\]
\[
\ba{c}
\ds{
\CX_s^{(2)}=\CX_{\bar{s}}^{(2)}= n ~.
}
\acc
\ea
\]
\vskip -5pt 
\noindent
These numbers enter into the central charge calculation via the
arguments of Rogers' dilogarithm functions 
\[
L(x)=-\half\int_0^x\!dy\lf[\frac{\ln y}{1{-}y}+\frac{\ln(1{-}y)}{y}\ri]~.
\]
The following sum rules were verified numerically, for all cases
up to $N{=}30\,$:
\[
c_{\Y}=\frac{6}{\pi^2}
\sum_{i,\alpha}  L \lf( {1 \over {1 +\Y_i^{(\alpha)}}} \ri)
= {2 (N-1) \over N+2} + \left\{ \begin{array}{ll}
$4$ & \mbox{for }N=4n{+}1\\
${5/2}$ & \mbox{for }N=4n{+}2\\
$2$ & \mbox{for }N=4n{+}3\\
 $1$ & \mbox{for }N=4n{+}4
\end{array} \right.
\]
\[
c_{\CX}=\frac{6}{\pi^2}
\sum_{i,\alpha}  L \lf( {1 \over {1 +\CX_i^{(\alpha)}}} \ri)
=  \left\{ \begin{array}{ll}
$4$ & \mbox{for }N=4n{+}1\\
${5/2}$ & \mbox{for }N=4n{+}2\\
$2$ & \mbox{for }N=4n{+}3\\
$1$ & \mbox{for }N=4n{+}4
\end{array} \right.
\]
\[
c_{\CZ}=\frac{6}{\pi^2}
\sum_{i,\alpha}  L \lf( {1 \over {1 +\CZ_i^{(\alpha)}}} \ri)
= \left\{ \begin{array}{ll}
$3$ & \mbox{for }N=4n{+}1\\
${3/2}$ & \mbox{for }N=4n{+}2\\
$1$ & \mbox{for }N=4n{+}3\\
$0$ & \mbox{for }N=4n{+}4
\end{array} \right.
\]
\smallskip\noindent
In fact, the identities found were more general than this: in the
spirit of ref.~\cite{GTa}, we checked that they also hold if the
constants $\CX$, $\Y$ and $\CZ$ are replaced by any
non-stationary solutions to the Y-systems of the last section, so long as 
an additional average over a full period is performed for each result. 
These periods are $2(N{+}2)$ for the $\Y$'s, $N$ for the $\CX$'s, and
$8$, $4$ and $8$ (when $\rho=1$,$2$ and $3$) for the $\CZ$'s.
 
\smallskip
\noindent 
The sum rules are not obviously
relevant when $N{=}4n{+}3$, since for these TBA systems the kernels
given in the text are not all symmetric.
However, at the expense 
of some additional complexity, the 
systems can be rewritten in a
symmetrical way,  after which the central charge calculation goes
through as usual. The forms given earlier  were selected as the
simplest versions which are also suitable for direct numerical
solution.

With this detail out of the way, the ultraviolet and infrared central
charges are
\eq
c_{UV}=c_{\Y}-c_{\CX} = { 2 (N{-}1) \over (N{+}2)}~;
\en
\eq
c_{IR}=c_{\CX}-c_{\CZ} = 1 ~.
\en
 
%
%
\section{Regular UV expansion coefficients}
This appendix tabulates fits of the numerical solutions to the TBA
equations at small $R\/$ to the expansions
\[
RE(R)+\frac{\pi c}{6}-({\rm bulk~piece})=
     2\pi\sum_{m=0}^{\infty}F_m\,(\MR)^{2{N{-}4\over N{+}2}m}~.
\]
The scale
$M$ was set equal to $1$, and $c$ to 
$2(N{-}1)/(N{+}2)$.
The bulk pieces, and some details of the numerical
procedures used, can be found in section~3 of the main text.
For completeness, coefficients for $N{=}8$, obtained simply by doubling 
those reported by Zamolodchikov for the tricritical Ising to Ising 
flow in ref.~\cite{Zd}, have also been included.
%
%
\begin{table}[hbt]
\widetable  
\begin{center}
\noindent\llap{\raisebox{-15pt}{$N{=}\,5\,$:\qquad}}
\begin{tabular}[t]{|c|l|l|}     \hline \hline
\rll 
{}~m~{} &~~~~$\wtilde F_m$ (massive)~~~     &~~~~$F_m$ (massless)~~~~\\
\hline
  0 & $~\m  1.98 \times 10^{-13}     $  &  $~  -3.86 \times 10^{-14} $ \\
\hline
  1 & $~   -7.78 \times 10^{-12}     $  &  $~\m 3.23 \times 10^{-12} $ \\
\hline
  2 & $~\m  0.0112031704~~~~         $  &  $~\m 0.0112031702~~~~     $ \\
\hline
  3 & $~   -3.37\times 10^{-09}      $  &  $~\m 1.35\times 10^{-09}  $ \\
\hline
  4 & $~\m  0.00038484               $  &  $~\m 0.00038477           $ \\
\hline
  5 & $~\m  0.019215                 $  &  $~  -0.019217             $ \\
\hline
  6 & $~\m  0.02056                  $  &  $~\m 0.02055              $ \\
\hline
  7 & $~\m  0.0004                   $  &  $~  -0.0006               $ \\
\hline\hline
\end{tabular}
\end{center}
\vskip -10pt
\end{table}   
%
%
%
%
\begin{table}[hbt]
\widetable  
\begin{center}
\vskip -5pt  
\noindent\llap{\raisebox{-15pt}{$N{=}\,6\,$:\qquad}}
\begin{tabular}[t]{|c|l|l|}     \hline \hline
\rll
{}~m~{} &~~~~$\wtilde F_m$ (massive)~~~     &~~~~$F_m$ (massless)~~~~\\
\hline   
  0 & $~\m 2.15\times 10^{-18} $ & $~\m 4.10\times 10^{-17}  $ \\
\hline
  1 & $~\m 1.01\times 10^{-13} $ & $~\m 1.22\times 10^{-13}  $ \\
\hline
  2 & $~\m 0.026663544239~~    $ & $~\m 0.026663544230~~     $ \\
\hline 
  3 & $~\m 0.0202428666        $ & $~  -0.0202428658         $ \\
\hline
  4 & $~\m 0.00088616          $ & $~\m 0.00088614           $ \\
\hline
  5 & $~  -0.0061087           $ & $~\m 0.0061090            $ \\
\hline
  6 & $~  -0.002800            $ & $~  -0.002803             $ \\
\hline
  7 & $~\m 0.00033             $ & $~  -0.00031              $ \\
\hline \hline
\end{tabular}
\end{center}
\vskip -10pt
\end{table}
%
\begin{table}
\widetable  
\begin{center}
\vskip -5pt
\noindent\llap{\raisebox{-15pt}{$N{=}\,7\,$:\qquad}}
\begin{tabular}[t]{|c|l|l|}     \hline \hline
\rll
{}~m~{} &~~~~$\wtilde F_m$ (massive)~~~     &~~~~$F_m$ (massless)~~~~\\
\hline
  0 & $~\m 3.98\times 10^{-15} $ & $~\m  8.02\times 10^{-16} $ \\
\hline
  1 & $~\m 2.42\times 10^{-14} $ & $~   -5.84\times 10^{-15} $ \\
\hline
  2 & $~\m 0.04302032556~~~    $ & $~\m  0.04302032557~~~    $ \\
\hline
  3 & $~\m 0.0128965993        $ & $~   -0.0128965995        $ \\
\hline
  4 & $~\m 0.000972018         $ & $~\m  0.000972029         $ \\
\hline
  5 & $~\m 0.00120282          $ & $~   -0.00120286          $ \\
\hline
  6 & $~\m 0.0000682           $ & $~\m  0.0000690           $ \\
\hline
  7 & $~  -0.0000085           $ & $~\m  0.0000077           $ \\
\hline  \hline
\end{tabular}
\end{center}
\vskip -10pt
\end{table}
%
\begin{table}
\widetable
\begin{center}
\vskip -5pt
\noindent\llap{\raisebox{-15pt}{$N{=}\,8\,$:\qquad}}
\begin{tabular}[t]{|c|l|l|}     \hline \hline
\rll
{}~m~{} &~~~~$\wtilde F_m$ (massive)~~~     &~~~~$F_m$ (massless)~~~~\\
\hline
  2 & $~\m 0.06589308648~~ $ & $~\m 0.06589308648~~ $ \\
\hline
  3 & $~  -0.02744520208    $ & $~  -0.02744520208    $ \\
\hline
  4 & $~\m 0.00073623262   $ & $~\m 0.00073623264   $ \\
\hline
  5 & $~\m 0.00102222      $ & $~  -0.00102216      $ \\
\hline
  6 & $~  -0.00016842      $ & $~  -0.00016851      $ \\ 
\hline
  7 & $~  -0.00004770      $ & $~\m 0.00004710      $ \\
\hline
  8 & $~  -0.0000268~~     $ & $~  -0.0000226~~     $ \\
\hline \hline
\end{tabular} 
\end{center}
\vskip -10pt
\end{table}
%
\begin{table}
\widetable  
\begin{center}
\vskip -5pt  
\noindent\llap{\raisebox{-15pt}{$N{=}\,9\,$:\qquad}}
\begin{tabular}[t]{|c|l|l|}     \hline \hline
\rll 
{}~m~{} &~~~~$\wtilde F_m$ (massive)~~~     &~~~~$F_m$ (massless)~~~~\\
\hline
0  & $~\m 7.27 \times 10^{-14} $ & $~\m 2.89 \times 10^{-13} $ \\
\hline   
1  & $~  -4.21 \times 10^{-12} $ & $~  -1.97 \times 10^{-11} $ \\
\hline   
2  & $~\m 0.124757776 $ & $~\m 0.124757776 $ \\
\hline   
3  & $~  -0.01159266 $  & $~\m 0.01159265  $ \\
\hline   
4  & $~\m 0.0004886 $   & $~\m 0.0004887 $ \\
\hline   
5  & $~\m 0.000574  $   & $~  -0.000575 $ \\
\hline   
6  & $~\m 0.000088  $   & $~\m 0.000089  $ \\
\hline   
7  & $~  -0.000028 $    & $~\m 0.000025 $ \\ 
\hline \hline 
\end{tabular} 
\end{center} 
\vskip-10pt 
\end{table} 
%
\begin{table}[h!]
\widetable  
\begin{center}
\vskip -5pt  
\noindent\llap{\raisebox{-15pt}{$N{=}\,10\,$:\qquad}}
\begin{tabular}[t]{|c|l|l|}     \hline \hline
\rll 
{}~m~{} &~~~~$\wtilde F_m$ (massive)~~~     &~~~~$F_m$ (massless)~~~~\\
\hline
0  & $~\m 1.17 \times 10^{-14}$ & $~  -2.94 \times 10^{-14}$ \\
\hline   
1  & $~  -8.58 \times 10^{-13}$ & $~\m 2.40 \times 10^{-12}$ \\
\hline   
2  & $~\m 0.0247185731 $ & $~\m 0.0247185730 $ \\
\hline   
3  & $~ -0.006211643   $ & $~\m 0.006211644  $ \\
\hline   
4  & $~\m  0.00031116  $ & $~\m 0.00031115   $ \\
\hline   
5  & $~\m  0.00030406  $ & $~  -0.00030402   $ \\
\hline   
6  & $~\m  0.0000402   $ & $~\m 0.0000401    $ \\
\hline   
7  & $~ -0.0000137     $ & $~\m 0.0000138    $ \\
\hline \hline
\end{tabular}
\end{center}
\vskip -10pt
\end{table}   
%
%
\clearpage
\newpage
%
%

\end{document}